\documentclass[prd,aps,floatfix,nofootinbib,twocolumn,tightenlines,showpacs]{revtex4}
\usepackage{dcolumn}% Align table columns on decimal point
\usepackage{latexsym}
\usepackage{graphicx}
\usepackage{bm}

\def\ttl#1{{``#1''}}
\def\svev#1{\left\langle #1\right\rangle}       % variable < >
\def\tr{{\rm tr}\,}
\def\Tr{{\rm Tr}\,}

\def\Re{{\rm Re\,}}

\long \def \blockcomment #1\endcomment{}

\def\Eq#1{Eq.~(\ref{#1})}
\def\tbeta{\tilde\beta}

\newcommand{\bee}{\begin{equation}}
\newcommand{\ee}{\end{equation}}
\newcommand{\beea}{\begin{eqnarray}}
\newcommand{\eea}{\end{eqnarray}}

\begin{document}
%%%%%%%%%%%%%%%%%%%%%%%%%%%%%%%%%%%%%%%%%%%%%%%%%%%%%%%%%%%%%%%%%%%%%%

\title{Near the sill of the conformal window:\\ gauge theories with fermions in two-index representations}
\author{Thomas DeGrand}
%\email{degrand@pizero.colorado.edu}
\affiliation{Department of Physics,
University of Colorado, Boulder, CO 80309, USA}
\author{Yigal Shamir}
%\email{shamir@post.tau.ac.il}
\author{Benjamin Svetitsky}
%\email{bqs@julian.tau.ac.il}
\affiliation{Raymond and Beverly Sackler School of Physics and Astronomy,
  Tel~Aviv University, 69978 Tel~Aviv, Israel\\ }

\begin{abstract}
We apply Schr\"odinger functional methods to two gauge theories with fermions in two-index representations: the SU(3) theory with $N_f=2$ adjoint fermions, and the SU(4) theory with
$N_f=6$ fermions in the two-index antisymmetric representation.
Each theory is believed to lie near the bottom of the conformal window for its respective representation.
In the SU(3) theory we find a small beta function in strong coupling but we cannot confirm or rule out an infrared fixed point.
In the SU(4) theory we find a hint of walking---a beta function that approaches the axis and then turns away from it.
In both theories the mass anomalous dimension remains small even at the strongest couplings, much like the theories with fermions in the two-index symmetric representation investigated earlier.
\end{abstract}

\pacs{11.15.Ha, 11.10.Hi, 12.60.Nz}
\maketitle

%%%%%%%%%%%%%%%%%%%%%%%%%%%%%%%%%%%%%%%%%%%%%%%%%%%%%%%%%%%%%%%%%%%%%
\section{Introduction}
%%%%%%%%%%%%%%%%%%%%%%%%%%%%%%%%%%%%%%%%%%%%%%%%%%%%%%%%%%%%%%%%%%%%%

The extension of lattice gauge methods to theories beyond QCD has been largely aimed at determining the infrared properties of these theories~\cite{Neil:2012cb,Giedt:2012it}.
For a given gauge group, one varies the number $N_f$ of fermion flavors to try to find the conformal window, the range of $N_f$ where the theory is scale invariant at large distances.
Below this window the theory confines and breaks global symmetries, much like QCD;
the most interesting range of $N_f$ is the borderline area \cite{Holdom:1981rm,Yamawaki:1985zg,Hill:2002ap}.
In a further departure from QCD, one can put the fermions in a color representation other than the
fundamental.
This opens a large arena for exploration~\cite{Sannino:2004qp,Hong:2004td,Dietrich:2006cm}.

In this paper we analyze two gauge theories:
the SU(3) gauge theory with $N_f=2$ Dirac fermions in the adjoint representation, and the SU(4) theory with $N_f=6$ Dirac fermions in the sextet, which is an antisymmetric tensor with two indices.
For the SU(3)/adjoint theory, $N_f=2$ is the only value that is interesting, in that the coefficients $b_1,b_2$ of the one- and two-loop terms in the beta function,
\bee
\beta(g^2)= -b_1\frac{g^4}{16\pi^2}-b_2\frac{g^6}{(16\pi^2)^2} + \cdots,
\ee
satisfy
\bee
b_1>0,\qquad b_2<0.
\label{b12}
\ee
The two-loop beta function thus possesses an infrared-stable fixed point (IRFP)~\cite{Caswell:1974gg,Banks:1981nn}, which invites non-perturbative confirmation.
As for the SU(4)/sextet theory, the condition (\ref{b12}) offers a wider range of $N_f$ for study
(see Table~\ref{table:bfun}).
Approximate solutions of the Schwinger--Dyson equations \cite{Appelquist:1988yc,Cohen:1988sq} indicate that the $N_f=6,7$ theories lie below the sill of the conformal window while $N_f=8$ lies just above (see Ref.~\cite{Dietrich:2006cm}).
Allowing that all three theories invite study, we chose to start with $N_f=6$ so as to approach the conformal window from below.\footnote{
As is well known, choosing an even number for $N_f$ allows a much simpler and less expensive algorithm for simulation than an odd number.
The $N_f=5$ theory also satisfies \Eq{b12}, but we have omitted it from the table.}

%%%%%%%%%%%%%%%%%%%%%%%%%%%%%%%%%%%%%%%%%%%%%%%%%%%%%%%%%%%%%%%%%%%%%
\begin{table}
\caption{Coefficients of the two-loop beta function for the SU(3)/adjoint
and SU(4)/sextet theories, and location of its zero $g_*^2$.
For comparison we list the quantities for borderline
theories with two-index symmetric representations.
\label{table:bfun}}
\begin{center}
\begin{ruledtabular}
\begin{tabular}{lcccd}
   &
  $N_f$ &
  $b_1$ &
  $b_2$&
  g_*^2
  \\
\hline
SU(3)/adjoint & 2 & 3         & $-90$        & 5.26 \\[2pt]
SU(4)/sextet  & 6 & $6\frac23$& $-38\frac23$ & 27.2 \\[2pt]
              & 7 & $5\frac13$& $-75\frac13$ & 11.2 \\[2pt]
              & 8 & 4         & $-112$       & 5.6  \\[5pt]
SU(2)/triplet & 2 & 2         & $-40$        & 7.9  \\[2pt]
SU(3)/sextet  & 2 & $4\frac13$& $-64\frac23$ & 10.6 \\[2pt]
SU(4)/decuplet& 2 & $6\frac23$& $-86\frac23$ & 12.1 \\
\end{tabular}
\end{ruledtabular}
\end{center}
\end{table}
%%%%%%%%%%%%%%%%%%%%%%%%%%%%%%%%%%%%%%%%%%%%%%%%%%%%%%%%%%%%%%%%%%%%%

We apply the method of the Schr\"odinger functional (SF) to calculate
the running coupling of the theories at hand, and thus their beta functions.
This method was developed
\cite{Luscher:1992an,Luscher:1993gh,Sint:1993un,Sint:1995ch,Jansen:1998mx,DellaMorte:2004bc}
to study small-$N_f$ QCD, whose coupling runs fast in evolving from short to long distance scales.
While we use the same definition of the running coupling, we analyze the results with methods that
we have found useful for conformal and near-conformal theories, where the coupling runs very slowly.
We developed these methods in the course of our work on three gauge theories that lie near the
bottom of the conformal window: the SU(2)~\cite{DeGrand:2011qd},
SU(3)~\cite{Shamir:2008pb,DeGrand:2010na,DeGrand:2012yq}, and SU(4)~\cite{DeGrand:2012qa} theories,
all with $N_f=2$ fermions in the respective two-index symmetric representations (2ISR) of color.%
\footnote{
  For other applications of the SF method to near-conformal gauge theories,
  see \cite{Appelquist:2007hu,Appelquist:2009ty,Hietanen:2009az,Bursa:2009we,Bursa:2010xn,Hayakawa:2010yn,Karavirta:2011zg,Hayakawa:2012gf,Voronov:2012qx,Rantaharju:2013gz}.
}

The present study takes us to new two-index representations---the adjoint and the antisymmetric.
Our SF analysis, even before extrapolation to the continuum limit, is inconclusive regarding the
existence of an IRFP in both theories we study.
The extrapolation of our data to the continuum is difficult.
The resulting error bars are on the same scale as the one-loop beta function, and so we cannot tell whether the beta function for
each theory crosses zero.
It is possible that it approaches zero and then runs off to negative values, behavior known as {\em walking}.
The SU(4)/sextet theory, in particular, shows a hint of this behavior, but with large error bars.

As in our work on the 2ISR models, we are able to give more precise results for $\gamma_m$, the anomalous dimension of the fermion mass, defined as usual through the scaling behavior of $\bar\psi\psi$.
This is calculated as a byproduct of the SF calculation \cite{Sint:1998iq,Capitani:1998mq,DellaMorte:2005kg,Bursa:2009we}.
We find, as in the other theories, that as $g^2$ is increased, $\gamma_m$ deviates downwards from the one-loop curve and levels off below 0.4 in both the SU(3) and the SU(4) theories.
This result is robust under continuum extrapolation.

The levelling-off of $\gamma_m(g)$ in these theories and in the 2ISR theories is a remarkable result.
No such behavior has ever appeared in perturbation theory.
As we have noted before, the existence of a global bound on $\gamma_m$ is evidently invariant under any redefinition $g\to g'(g)$, that is, it is entirely scheme-independent.

The SU(3) lattice gauge theory with two adjoint fermions has attracted attention as an extension of QCD in which the dynamical scales of confinement and of chiral symmetry breaking might be separated.
Following early quenched \cite{Kogut:1984sb} and unquenched \cite{Gerstenmayer:1989qw} studies,
Karsch and L\"utgemeier \cite{Karsch:1998qj} carried out an extensive study of the $N_f=2$ theory with staggered fermions.
They found clear evidence for separate finite-temperature phase transitions.
The nature of the chiral phase transition, which should follow the scheme \cite{Peskin:1980gc} $\textrm{SU}(4)\to\textrm{SO}(4)$, was investigated in Refs.~\cite{Basile:2004wa,Engels:2005te}.
The theory has also served as a laboratory for studying monopole condensation \cite{Cossu:2008wh} and finite-size phase transitions \cite{Unsal:2007jx,Cossu:2009sq}.

The finite-temperature transitions found in the above work would seem to rule out IR conformality in the SU(3)/adjoint theory.
After all, an IR conformal theory would have no scale from which one could construct a zero-temperature chiral condensate, a string tension, or a transition temperature.
The evidence offered so far, however, is inconclusive.
The results cited above were obtained in studies on finite-temperature lattices with only one value of $n_\tau$, the number of sites in the Euclidean time direction.
When $n_\tau=4$, say, the lattice spacing itself sets a scale for the temperature so that a transition occurs at some bare coupling $g_0^*$.
A true test of confinement vs.~conformality requires varying $n_\tau$ to see the behavior of $g_0^*(n_\tau)$.
This will determine whether,  in the continuum limit, the transition temperature reaches a finite limit or tends to zero.
Such a program has been attempted for the SU(3)/triplet theory with various $N_f$ \cite{Deuzeman:2008sc,Miura:2011mc,Cheng:2011ic,Deuzeman:2012ee,Miura:2012zqa} and for the SU(3)/sextet theory \cite{Kogut:2010cz,Kogut:2011ty,Kogut:2011bd,Sinclair:2012fa}, and it is fraught with difficulties.\footnote{
These studies used the staggered prescription for the fermions; the finite-temperature transition in the SU(3)/sextet theory was observed with Wilson fermions in Ref.~\cite{DeGrand:2008kx}.}

The SU(4)/sextet theory has not been studied on the lattice before.
It stands out in Table~\ref{table:bfun}.
For $N_f=6$, the zero of the two-loop beta function occurs at $g_*^2\simeq27.2$.
This is a much stronger coupling than in the other borderline theories listed in the table.
In the 2ISR theories \cite{DeGrand:2011qd,Shamir:2008pb,DeGrand:2010na,DeGrand:2012yq,DeGrand:2012qa} as well as in the SU(3)/adjoint theory (see below), we found that the nonperturbative beta function follows the two-loop form fairly closely out to its zero.\footnote{
We were able to confirm the zero at high significance in the SU(2) theory \cite{DeGrand:2011qd} but not in the other theories.}
Clearly the two-loop beta function cannot be trusted out to $g^2=27$, and in fact we will show below that the calculated beta function deviates and approaches zero at a much weaker coupling.

We review the choice of lattice action and describe our simulations in Sec.~\ref{sec:action}.
We present the analysis of the running coupling and the beta function in both theories in Sec.~\ref{sec:coupling},
and the mass anomalous dimension in Sec.~\ref{sec:gamma}.
We conclude with a summary of our results and a discussion of the difficulties encountered.

%%%%%%%%%%%%%%%%%%%%%%%%%%%%%%%%%%%%%%%%%%%%%%%%%%%%%%%%%%%%%%%%%%%%%
\section{Lattice action, phase diagram, and ensembles\label{sec:action}}
%%%%%%%%%%%%%%%%%%%%%%%%%%%%%%%%%%%%%%%%%%%%%%%%%%%%%%%%%%%%%%%%%%%%%

Our fermion action $\bar\psi D_F \psi$ is the conventional Wilson action,
supplemented by a clover term~\cite{Sheikholeslami:1985ij} with coefficient
$c_{\text{SW}}=1$~\cite{Shamir:2010cq}.
The gauge links in the fermion action are fat link variables $V_\mu(x)$.
The fat links are the normalized
hypercubic (nHYP) links of Refs.~\cite{Hasenfratz:2001hp,Hasenfratz:2007rf}
with weighting parameters $\alpha_1=0.75$, $\alpha_2=0.6$,
$\alpha_3=0.3$, subsequently promoted to the fermions'
representation.

As we found in our previous work~\cite{DeGrand:2012qa,DeGrand:2012yq}, it is useful to generalize the pure gauge
part of the action beyond the usual plaquette term to include a term
built out of fat links.
Thus,
\begin{eqnarray}
S_G &=&-
\frac{\beta}{2N} \sum_{\mu\ne\nu} \Re \Tr U_\mu(x) U_\nu(x+\hat\mu)
U_\mu^\dagger(x+\hat\nu) U_\nu^\dagger(x)
\nonumber\\
& &- \frac{\beta'}{2d_f} \sum_{\mu\ne\nu} \Re \Tr V_\mu(x) V_\nu(x+\hat\mu)
V_\mu^\dagger(x+\hat\nu) V_\nu^\dagger(x).
\nonumber\\
\label{eq:gaugeaction}
\end{eqnarray}
$N=3,4$ is the number of colors while $d_f$ ($=8,6$, respectively) is the dimension of the fermion
representation.

\begin{figure}
\begin{center}
\includegraphics[width=.95\columnwidth,clip]{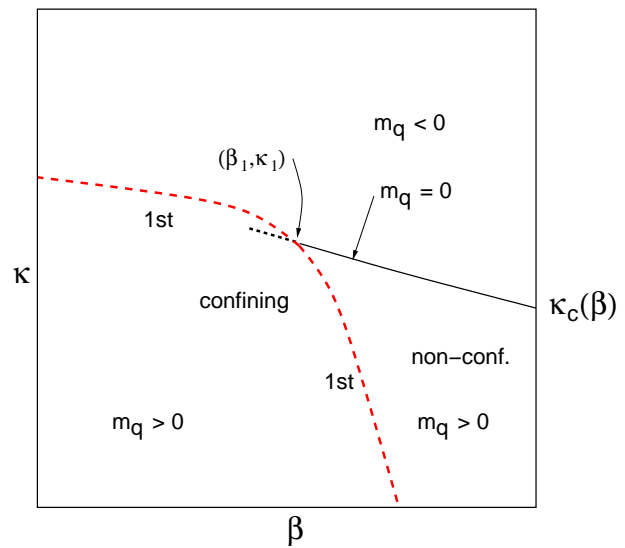}
\end{center}
\caption{
  Presumed phase diagram for both the SU(3)/adjoint and SU(4)/sextet theories,
for $|\beta'|$ not too large.
The first-order phase boundary, as well as the $\kappa_c(\beta)$ curve, shifts with $\beta'$;
the SF coupling $g$ is determined along the $\kappa_c$ curve, and it, too,
depends on $\beta'$.
\label{fig:sketch}}
\end{figure}

The reason for adding the $\beta'$ term can be found in the phase diagram sketched in Fig.~\ref{fig:sketch} \cite{Iwasaki:1991mr,Iwasaki:2003de,Nagai:2009ip}.
We verified this phase diagram in the SU(3)/sextet theory \cite{DeGrand:2010na,DeGrand:2008kx} and  in the SU(2)/triplet theory \cite{DeGrand:2011qd} (where part of the phase boundary is a second-order transition).%
\footnote{For more discussion see
Refs.~\cite{Svetitsky:2010zd,Svetitsky:2013px}.}
For the two theories at hand,
we discovered the same structure in the course of determining the $\kappa_c(\beta)$ curve.
Our SF calculations are, in principle, carried out on the $\kappa_c(\beta)$ line, where $m_q=0$, to the right of the point marked $(\beta_1,\kappa_1)$.
Our goal is to reach as strong a running coupling as possible, by pushing to strong bare couplings.
At strong coupling, however, we encounter either
the phase transition shown or impossibly low acceptance due to roughness of the
typical gauge configuration.
Adjusting $\beta'$ can help push these limitations off towards stronger
coupling.
An exploration of the $(\beta,\beta')$ plane similar to that described in
Ref.~\cite{DeGrand:2012qa} led us to set $\beta'=0.5$ in the SU(4) theory.
In the SU(3) theory, on the other hand, we found no advantage in adjusting
$\beta'$ away from zero.
Any other choice {\em decreased\/} the range of accessible couplings.
See the Appendix for more information.
%%%%%%%%%%%%%%%%%%%%%%%%%%%%%%%%%%%%%%%%%%%%%%%%%%%%%%%%%%%%%%%%%%%%%
%%%%%%%%%%%%%%%%%%%%%%%%%%%%%%%%%%%%%%%%%%%%%%%%%%%%%%%%%%%%%%%%%%%%%

We determine the critical hopping
parameter $\kappa_c = \kappa_c(\beta)$ by setting to zero the quark mass $m_q$,
as defined by the unimproved axial Ward identity.
$m_q$ is of course volume-dependent on small lattices.
Ideally, we would like to fix $\kappa_c$ so that $m_q\to0$ in the infinite-volume limit.
For clear practical reasons, we instead do the determination in relatively short runs on lattices of size $L=12a$.
In our work on the SU(2)/adjoint theory \cite{DeGrand:2011qd}, we addressed the concern that an extrapolation of $m_q$ to infinite volume
might show that the $L\to\infty$ limit is far
from massless.
The problem is potentially serious only at the strongest couplings, and we showed that, even there, adjustment of $\kappa$ to make $m_q$ acceptably small does not affect the results for the $\beta$ function or the anomalous dimension $\gamma_m$.

\begin{figure*}
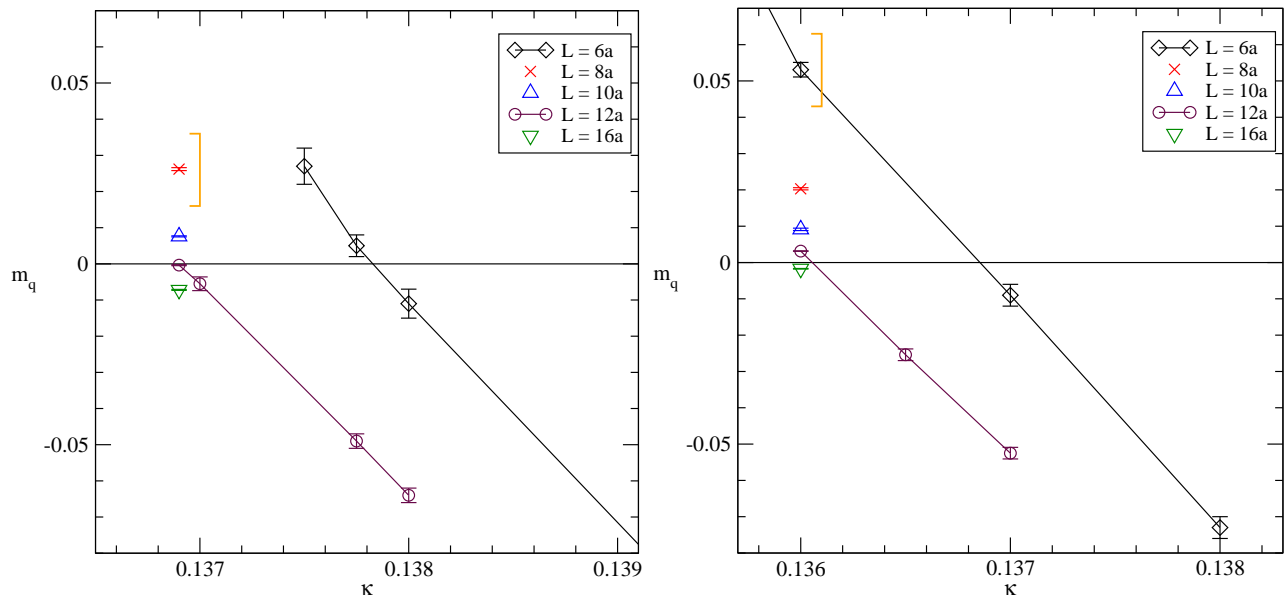

\begin{center}
\includegraphics[width=.47\textwidth,clip]{mq_38.eps}\
\includegraphics[width=.47\textwidth,clip]{mq_39.eps}
\end{center}
\caption{
$m_q(\kappa)$, determined from the axial Ward identity, for the strongest
couplings studied
in the SU(3)/adjoint theory.
Left: $\beta=3.8$.  Right: $\beta=3.9$.
The square brackets indicate measurements in a metastable state.
\label{fig:mq}}
\end{figure*}

In the SU(3)/adjoint theory, on the other hand, the determination of $\kappa_c$ runs into trouble
at the strongest couplings, $\beta=3.8$ and~3.9, and this time the problem lies in the smallest lattices.
We refer again to Fig.~\ref{fig:sketch}.
Both the first-order phase boundary and the $\kappa_c(\beta)$ curve shift with
volume.
We show in Fig.~\ref{fig:mq} the data for $m_q(\kappa)$ for the various volumes
at the two strongest bare couplings, $\beta=3.8$ and~3.9.
%%%%%%%%%%%%%%%%%%%%%%%%%%%%%%%%%%%%%%%%%%%%%%%%%%%%%%%%%%%%%%%%%%%%%
%%%%%%%%%%%%%%%%%%%%%%%%%%%%%%%%%%%%%%%%%%%%%%%%%%%%%%%%%%%%%%%%%%%%%
For each coupling we fix $\kappa_c$ by demanding $m_q=0$ for $L=12a$.
At $\beta=3.8$, this fixes $\kappa_c=0.1369$.
As can be seen in the figure, the values of $m_q$ at this $\kappa$ for $L=10a$
and $L=8a$ are nonzero and positive.
This would hold as well for $L=6a$, except that for $L=6a$ the phase boundary
in Fig.~\ref{fig:sketch} has moved up past $\kappa=0.1369$ so that we find
ourselves in the confining phase.
It is impossible to simulate for $L=6a$ at the $\kappa_c$ determined at
$L=12a$.
It is worth noting that there is still a value of $\kappa$ at which $m_q$ crosses zero for $L=6a$; like the phase boundary, it has shifted upwards.
The bottom line is that we are prevented from simulating on the $L=6a$ lattice at $\kappa_c$.

For $\beta=3.9$ the situation is the same, except that for $L=6a$ at $\kappa=
\kappa_c=0.1360$ we succeeded in simulating in a short run in the metastable state that is non-confining.
(This is the origin of the bracketed point in the figure on the right.)
The lifetime of the metastable state, however, was too short to make it useful for an SF measurement.
Going back to $\beta=3.8$, we found in fact that for $L=8a$ as well, the
non-confining state is metastable at $\kappa_c$.
In this case, however, we were able to run a very long simulation and thus
to make a useful determination of the SF observables.

We stress that the metastability issue on the small volumes at $\beta=3.8$
and~3.9 is distinct from what happens to the left of $(\beta_1,\kappa_1)$
in Fig.~\ref{fig:sketch}.
The strong-coupling part of the phase boundary is a place where $m_q$ flips
sign discontinuously, and there is no equilibrium measurement that will
give $m_q=0$ for any volume~\cite{DeGrand:2011qd}.
At $\beta=3.8$ and~3.9, on the other hand, each lattice size allows a value of
$\kappa$ where $m_q=0$.
The fact that this $\kappa$ shifts with $L$ does not pose a special problem;
the shift in the phase boundary {\em does\/} pose a practical problem in preventing
simulation at a given $\kappa$ for small volumes.
We overcame this in the case of
$(\beta=3.8,L=8a)$, however, and so we make use of the data here even though the
state is metastable.%
\footnote{In fact, we did not determine finally which is the stable state for $L=8a$
and which is the metastable.
It is possible that the confined state will tunnel back to the nonconfined in short order,
but simulation of the confined state is very difficult due to poor acceptance so we did
not resolve this question.}

Our tables for the SU(3)/adjoint theory are thus missing entries for $L=6a$ at the two strongest couplings.
This problem did not appear in the SU(4)/sextet theory.
Moreover, we did not run simulations for $L=16a$ at the weakest coupling in either theory.

As before, we employed the hybrid Monte Carlo (HMC) algorithm in our
simulations.  The molecular dynamics integration was accelerated with an
additional heavy pseudo-fermion field as suggested by
Hasenbusch~\cite{Hasenbusch:2001ne}, multiple time
scales~\cite{Urbach:2005ji}, and a second-order Omelyan
integrator~\cite{Takaishi:2005tz}.
The ensembles generated for the two theories are listed in
Tables~\ref{table:su3runs} and~\ref{table:su4runs}.
For the SU(3) theory, each ensemble for $L\ge10a$ was divided
into four streams, with the requirement that the observables from
the four streams be consistent to a low $\chi^2$.
For the SU(4) theory this was done for all the ensembles, including
$L=6a,8a$.

The SU(3) theory was particularly difficult to simulate in its strong coupling region, requiring short trajectories and producing long autocorrelation times, which in turn resulted in slow convergence of the separate streams.
For several values of $(\beta,L)$, we were unable to satisfy our consistency test of $\chi^2<6$ for 3 degrees of freedom, in one observable or another, among the four streams.
In most of these cases, however, we saw a steady improvement with the length of the run.
Moreover,
we found that the high $\chi^2$ was caused by one stream out of the four; dropping this stream in favor of the majority resulted in a change of the mean that was less than $1\sigma$.
We decided therefore to deem these results statistically consistent.
The only exception arose at $\beta=3.8$ for $L=16a$, the largest volume at the strongest coupling.
Here an outlying stream resulted in $\chi^2=16/3$~dof in the result for $Z_P$, with no sign of improvement as the streams grew longer.
We were left with no choice but to omit this stream from the final average, resulting in a shift by $2.5\sigma$.
This one result for $Z_P$ is thus less reliable than the others and we mark it so in the following.

The tables show that the SU(4) theory reached reasonable error bars with much shorter simulations.
There were no special problems with $\chi^2$ among the streams once they had become long enough.

%%%%%%%%%%%%%%%%%%%%%%%%%%%%%%%%%%%%%%%%%%%%%%%%%%%%%%%%%%%%%%%%%%%%%
\begin{table}	% created for LM 26.5.2013.  Should be rechecked.
\caption{Ensembles generated at the bare couplings
  $(\beta, \kappa_c)$, on lattice sizes $L$, for the SU(3)/adjoint theory.
For this theory no fat-plaquette term was added to the action.
  Listed are the total number of trajectories for all streams at given
$(\beta,L)$,
  the trajectory length, and the HMC acceptance.}
\begin{center}
\begin{ruledtabular}
\begin{tabular}{ccrrcc}
  $\beta$ &
  $\kappa_c $ &
  $L/a$  &
  trajectories&
  trajectory&
  acceptance\\
&&&(thousands)&length&\\
\hline
3.8 & 0.1369  &  8 &  21 & 1.0 & 0.46 \\
            & & 10 &  62 & 0.5 & 0.65 \\
            & & 12 &  80 & 0.5 & 0.64 \\
            & & 16 &  88 & 0.4 & 0.43 \\[2pt]
3.9 & 0.136   &  8 &  37 & 1.0 & 0.68 \\
            & & 10 &  76 & 0.5 & 0.77 \\
            & & 12 & 133 & 0.5 & 0.72 \\
            & & 16 & 100 & 0.4 & 0.57 \\[2pt]
4.1 & 0.13454 &  6 &  26 & 1.0 & 0.85 \\
            & &  8 &  18 & 1.0 & 0.67 \\
            & & 10 &  38 & 0.5 & 0.87 \\
            & & 12 &  48 & 0.5 & 0.84\\
            & & 16 &  26 & 0.5 & 0.70 \\[2pt]
4.5 & 0.13172 &  6 &  16 & 1.0 & 0.99 \\
            & &  8 &   9 & 1.0 & 0.97 \\
            & & 10 &  13 & 1.0 & 0.94 \\
            & & 12 &  19 & 1.0 & 0.88\\
            & & 16 &  10 & 1.0 & 0.81 \\[2pt]
5.0 & 0.1295  &  6 &  17 & 1.0 & 0.99 \\
            & &  8 &   8 & 1.0 & 0.99 \\
            & & 10 &  13 & 1.0 & 0.99 \\
            & & 12 &  32 & 1.0 & 0.97\\
\end{tabular}
\end{ruledtabular}
\end{center}
\label{table:su3runs}
\end{table}
%%%%%%%%%%%%%%%%%%%%%%%%%%%%%%%%%%%%%%%%%%%%%%%%%%%%%%%%%%%%%%%%%%%%%
%%%%%%%%%%%%%%%%%%%%%%%%%%%%%%%%%%%%%%%%%%%%%%%%%%%%%%%%%%%%%%%%%%%%%
\begin{table}	% created for LM 27.5.2013.  Ensemble data completed by TD 1.7.2013.
\caption{Ensembles generated at the bare couplings
  $(\beta, \kappa_c)$, on lattice sizes $L$, for the SU(4)/sextet theory.
For this theory a fat-plaquette term was added to the action with coefficient
$\beta'=0.5$.
Columns as in Table~\ref{table:su3runs}.}
\begin{center}
\begin{ruledtabular}
\begin{tabular}{ccrrcc}
  $\beta$ &
  $\kappa_c $ &
  $L/a$  &
  trajectories&
  trajectory&
  acceptance\\
&&&(thousands)&length&\\
\hline
 5.5&0.13398&  6 & 8 & 1.0    & 0.74      \\
    &       &  8 & 8 & 0.5     & 0.79      \\
    &       & 10 &16 & 0.5 & 0.77 \\
    &       & 12 &48 & 0.5 & 0.57 \\
    &       & 16 &17 & 0.5 & 0.38 \\[2pt]
 6.0&0.13315&  6 & 8 & 1.0     & 0.64      \\
    &       &  8 & 8 & 1.0     & 0.48      \\
    &       & 10 & 8 & 0.5     & 0.45     \\
    &       & 12 &10 & 0.5 & 0.73 \\
    &       & 16 &16 & 0.5 & 0.57 \\[2pt]
 7.0&0.13120&  6 & 8 &1.0     &  0.92     \\
    &       &  8 & 8 &1.0     &  0.84    \\
    &       & 10 & 8 &1.0     &  0.67    \\
    &       & 12 &16 &1.0     &  0.47    \\
    &       & 16 &11 & 0.5 & 0.82 \\[2pt]
 8.0&0.12933&  6 & 8 & 1.0    & 0.98     \\
    &       &  8 & 8 & 1.0    & 0.96     \\
    &       & 10 & 8 & 1.0    & 0.93     \\
    &       & 12 &16 & 1.0    & 0.65     \\
    &       & 16 &16 & 0.5 & 0.86 \\[2pt]
10.0&0.12702&  6 & 8 & 1.0    &  0.99     \\
    &       &  8 & 8 & 1.0    &   0.99    \\
    &       & 10 & 8 & 1.0    &   0.99    \\
    &       & 12 & 8 & 1.0    &    0.97   \\
\end{tabular}
\end{ruledtabular}
\end{center}
\label{table:su4runs}
\end{table}
%%%%%%%%%%%%%%%%%%%%%%%%%%%%%%%%%%%%%%%%%%%%%%%%%%%%%%%%%%%%%%%%%%%%%%
%%%%%%%%%%%%%%%%%%%%%%%%%%%%%%%%%%%%%%%%%%%%%%%%%%%%%%%%%%%%%%%%%%%%%
\section{The running coupling and the beta function\label{sec:coupling}}
%%%%%%%%%%%%%%%%%%%%%%%%%%%%%%%%%%%%%%%%%%%%%%%%%%%%%%%%%%%%%%%%%%%%%

We compute the running coupling in the SF method exactly as described
in our previous papers.
We impose Dirichlet boundary
conditions at the time slices $t=0,L,$ and measure the response of the
quantum effective action.  The coupling emerges from a measurement of
the derivative of the action with respect to a parameter $\eta$ in the
boundary gauge field,
\begin{equation}
 \frac{K}{g^2(L)}  =
  \left.\svev{\frac{\partial S_{G}}{\partial\eta}
  -\tr \left( \frac{1}{D_F^\dagger}\;
        \frac{\partial (D_F^\dagger D_F)}{\partial\eta}\;
            \frac{1}{D_F} \right)}\right|_{\eta=0}  .
            \label{deta}
\end{equation}
The boundary conditions we use for each theory are copied from other theories
with the same gauge group.
For the SU(3)/adjoint model, see our paper on the SU(3)/sextet theory~\cite{DeGrand:2010na};
for the SU(4)/sextet theory see our paper on the SU(4)/decuplet theory~\cite{DeGrand:2012qa}.
The constant $K=12\pi$ emerges directly from the classical continuum action.

\begin{figure}
\begin{center}
\includegraphics[width=.95\columnwidth,clip]{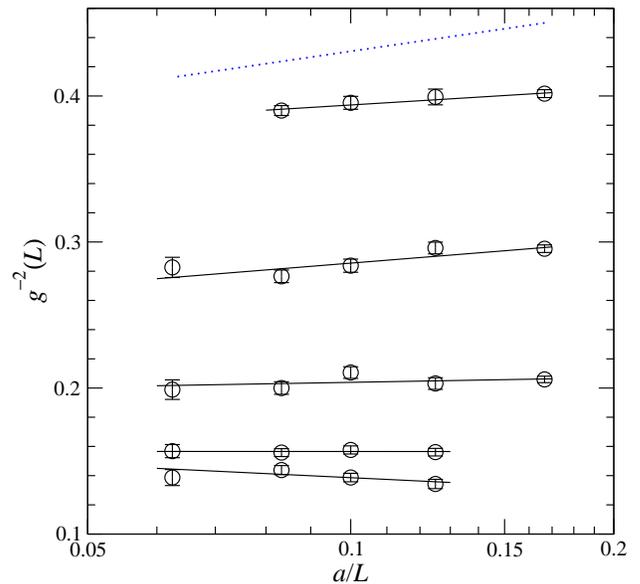}
\end{center}
\caption{
  Running coupling
  $1/g^2$ vs.~$a/L$ in the SU(3)/adjoint theory (Table \ref{table:su3couplings}).
  Top to bottom: $\beta=5.0,$ 4.5, 4.1, 3.9, 3.8.
  The straight lines are linear fits~[\Eq{linfit}]
  to each set of points at given $\beta$;
  the slope gives the beta function.
  The dotted line shows the expected slope from one-loop running.
\label{fig:su3couplings}}
\end{figure}

\begin{figure}
\begin{center}
\includegraphics[width=.95\columnwidth,clip]{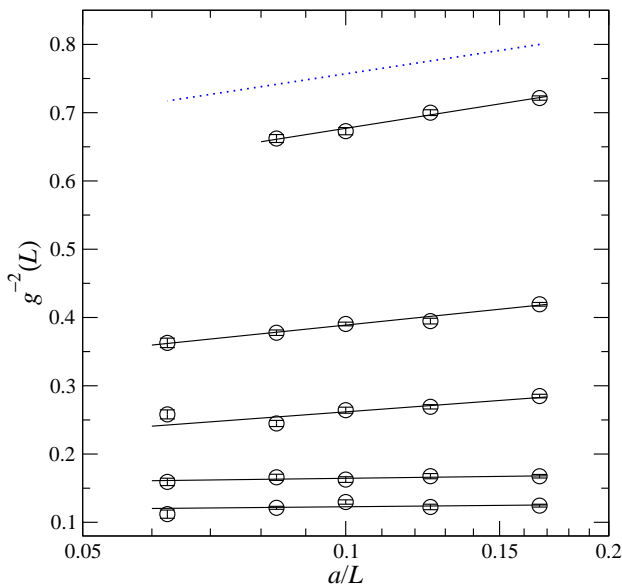}
\end{center}
\caption{Same as Fig.~\ref{fig:su3couplings}, but for
  the SU(4)/sextet theory (Table \ref{table:su4couplings}).
  Top to bottom: $\beta=10.0,$ 8.0, 7.0, 6.0, 5.5.
\label{fig:su4couplings}}
\end{figure}

We list the calculated running couplings for the SU(3) theory in Table~\ref{table:su3couplings} and for the SU(4) theory in Table~\ref{table:su4couplings};
they are plotted in Figs.~\ref{fig:su3couplings} and~\ref{fig:su4couplings},
respectively.
%%%%%%%%%%%%%%%%%%%%%%%%%%%%%%%%%%%%%%%%%%%%%%%%%%%%%%%%%%%%%%%%%%%%%%
\begin{table*}
\caption{Running coupling measured in the SU(3)/adjoint theory.}
\begin{center}
\begin{ruledtabular}
\begin{tabular}{ccllll}
$\beta$ &  \multicolumn{5}{c}{$1/g^2$}\\
\cline{2-6}
    & $L=6a$     & $L=8a$     & $L=10a$    & $L=12a$    & $L=16a$  \\
\hline
3.8 &  \hfil--   & 0.1343(32) & 0.1387(29) & 0.1438(31) & 0.1387(55) \\
3.9 &  \hfil--   & 0.1561(26) & 0.1576(28) & 0.1558(27) & 0.1568(45) \\
4.1 & 0.2059(22) & 0.2031(40) & 0.2106(40) & 0.2000(43) & 0.1989(67) \\
4.5 & 0.2954(26) & 0.2959(40) & 0.2838(45) & 0.2765(43) & 0.2826(69) \\
5.0 & 0.4016(27) & 0.3993(54) & 0.3953(44) & 0.3900(34) &  \hfil--   \\
\end{tabular}
\end{ruledtabular}
\end{center}
\label{table:su3couplings}
\end{table*}
%%%%%%%%%%%%%%%%%%%%%%%%%%%%%%%%%%%%%%%%%%%%%%%%%%%%%%%%%%%%%%%%%%%%%
%%%%%%%%%%%%%%%%%%%%%%%%%%%%%%%%%%%%%%%%%%%%%%%%%%%%%%%%%%%%%%%%%%%%%
%%%%%%%%%%%%%%%%%%%%%%%%%%%%%%%%%%%%%%%%%%%%%%%%%%%%%%%%%%%%%%%%%%%%%
%%%%%%%%%%%%%%%%%%%%%%%%%%%%%%%%%%%%%%%%%%%%%%%%%%%%%%%%%%%%%%%%%%%%%%
\begin{table*}
\caption{Running coupling measured in the SU(4)/sextet theory.}
\begin{center}
\begin{ruledtabular}
\begin{tabular}{ccllll}
$\beta$ &  \multicolumn{5}{c}{$1/g^2$}\\
\cline{2-6}
    & $L=6a$     & $L=8a$     & $L=10a$    & $L=12a$    & $L=16a$  \\
\hline
 5.5 & 0.1244(21) & 0.1225(40) & 0.1297(32) & 0.1213(22) & 0.1120(60) \\
 6.0 & 0.1675(26) & 0.1676(38) & 0.1626(42) & 0.1659(45) & 0.1592(54) \\
 7.0 & 0.2849(27) & 0.2692(32) & 0.2642(45) & 0.2448(45) & 0.2581(66) \\
 8.0 & 0.4193(27) & 0.3947(42) & 0.3905(27) & 0.3777(40) & 0.3628(69) \\
10.0 & 0.7214(32) & 0.7000(42) & 0.6729(53) & 0.6621(58) &  \hfil--   \\
\end{tabular}
\end{ruledtabular}
\end{center}
\label{table:su4couplings}
\end{table*}
%%%%%%%%%%%%%%%%%%%%%%%%%%%%%%%%%%%%%%%%%%%%%%%%%%%%%%%%%%%%%%%%%%%%%
%%%%%%%%%%%%%%%%%%%%%%%%%%%%%%%%%%%%%%%%%%%%%%%%%%%%%%%%%%%%%%%%%%%%%
%%%%%%%%%%%%%%%%%%%%%%%%%%%%%%%%%%%%%%%%%%%%%%%%%%%%%%%%%%%%%%%%%%%%%

We define the beta function $\tbeta(u)$ for $u\equiv1/g^2$ as
\bee
  \tbeta(u) \equiv \frac{d(1/g^2)}{d\log L}
  = 2\beta(g^2)/g^4 = 2u^2 \beta(1/u).
\label{invbeta}
\ee
As discussed in Ref.~\cite{DeGrand:2011qd}, the slow running of the
coupling suggests extracting the beta function at each $(\beta,\kappa_c)$ from
a linear fit of the inverse coupling
\bee
u(L)=c_0+c_1 \log \frac L{8a}\ .
\label{linfit}
\ee
With this parametrization, $c_0$ gives the inverse coupling $u(L=8a)$,
while $c_1$ is an estimate for the beta function $\tbeta$ at this coupling.

\begin{figure*}
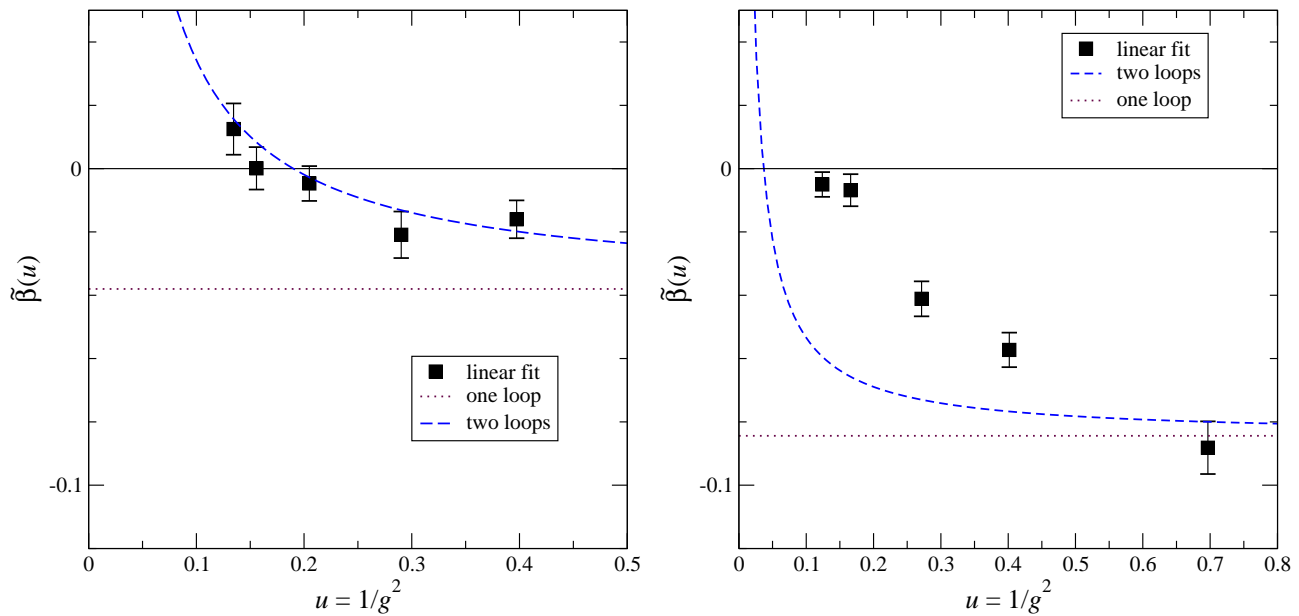

\begin{center}
\includegraphics[width=.47\textwidth,clip]{SU3adjbeta_scaled_linear_only.eps}
\
\includegraphics[width=.47\textwidth,clip]{SU4as2beta_linear_only.eps}
\end{center}
\caption{
Beta function $\tbeta(u)$ of the SU(3)/adjoint theory (left)
and the SU(4)/sextet theory (right), plotted as a function of $u(L=8a)$.
Results are extracted from the
linear fits~(\ref{linfit}), as shown in Figs.~\ref{fig:su3couplings} and~\ref{fig:su4couplings}, respectively.
Plotted curves are the one-loop (dotted line)
and two-loop (dashed line) beta functions.
No correction has been made for discretization errors.
\label{fig:beta}}
\end{figure*}

For a first look, we fit the data points for all $L$ to extract the
slopes at the given bare parameters, ignoring discretization errors that must be inherent in the smallest lattices.
These fits are shown in Figs.~\ref{fig:su3couplings} and~\ref{fig:su4couplings}.
Values of the beta function $\tbeta(u)$ obtained from these fits
are plotted as a function of $u(L=8a)$ in Fig.~\ref{fig:beta}.
%%%%%%%%%%%%%%%%%%%%%%%%%%%%%%%%%%%%%%%%%%%%%%%%%%%%%%%%%%%%%%%%%%%%%
%%%%%%%%%%%%%%%%%%%%%%%%%%%%%%%%%%%%%%%%%%%%%%%%%%%%%%%%%%%%%%%%%%%%%
Also shown are the one- and two-loop approximations from the expansion
(see Table~\ref{table:bfun})
\begin{equation}
  \tbeta(u) = -\frac{2b_1}{16\pi^2}
  -\frac{2b_2}{(16\pi^2)^2}\frac1u +\cdots .
\label{btilde}
\end{equation}

The plotted points for the SU(3) theory follow the two-loop curve closely, including its zero crossing.
This would imply an IRFP, but at low significance since the leftmost point is but $1.5\sigma$ above zero.
(We will see also that continuum extrapolation drives the point negative.)
In the case of the SU(4) theory, we see a large deviation from the two-loop curve in strong coupling, even tending towards a zero crossing but not quite getting there.
In both cases, one might be tempted to draw a smooth curve that crosses zero, but one could also draw a curve that approaches zero and then falls away, which is just the conjectured behavior for walking.

In Ref.~\cite{DeGrand:2012yq} we introduced a method for extrapolating
lattice results to the continuum limit when a theory runs slowly.
The key observation is that when a theory is almost conformal, the
finite lattice corrections will not depend separately on $a$ and on $L$ but
only on the ratio $(a/L)$.
Then successive elimination of the lattices with coarsest lattice spacing
$a$ is equivalent to dropping the smallest lattice sizes $L$.
We calculated $\tbeta(u)$ above by linear fits [\Eq{linfit}] to $1/g^2$ measured on
lattices of size $L_1<L_2<\ldots<L_N$.
The results for this first fit are the coefficients
$c_0\equiv c_0^{(1)}$ and $c_1\equiv c_1^{(1)}$.
We can obtain results closer to the continuum limit by dropping the smallest
lattice $L_1$ from consideration, whereupon a linear fit gives
$c_0^{(2)},c_1^{(2)}$.
Dropping the two smallest lattices gives $c_0^{(3)},c_1^{(3)}$,
and so forth.
Each $c_1^{(n)}$ is an approximant to $\tbeta(u)$ associated with $L=L_n$, the smallest lattice kept.
We can then extrapolate to $(a/L)=0$ either linearly,
\begin{equation}
c_1^{(n)}=\tbeta(u)+C\frac aL,
\label{linex}
\end{equation}
or quadratically,
\begin{equation}
c_1^{(n)}=\tbeta(u)+C\left(\frac aL\right)^2.
\label{quadex}
\end{equation}
Each extrapolation formula should be considered a model, since perturbative estimates of lattice error are inapplicable in the strong-coupling regime where we work.
The extrapolations take into account the fact that the results $c_1^{(n)}$ of the successive fits are correlated~\cite{DeGrand:2012yq}.
For graphs illustrating the procedure, see the Appendix.

\begin{figure*}
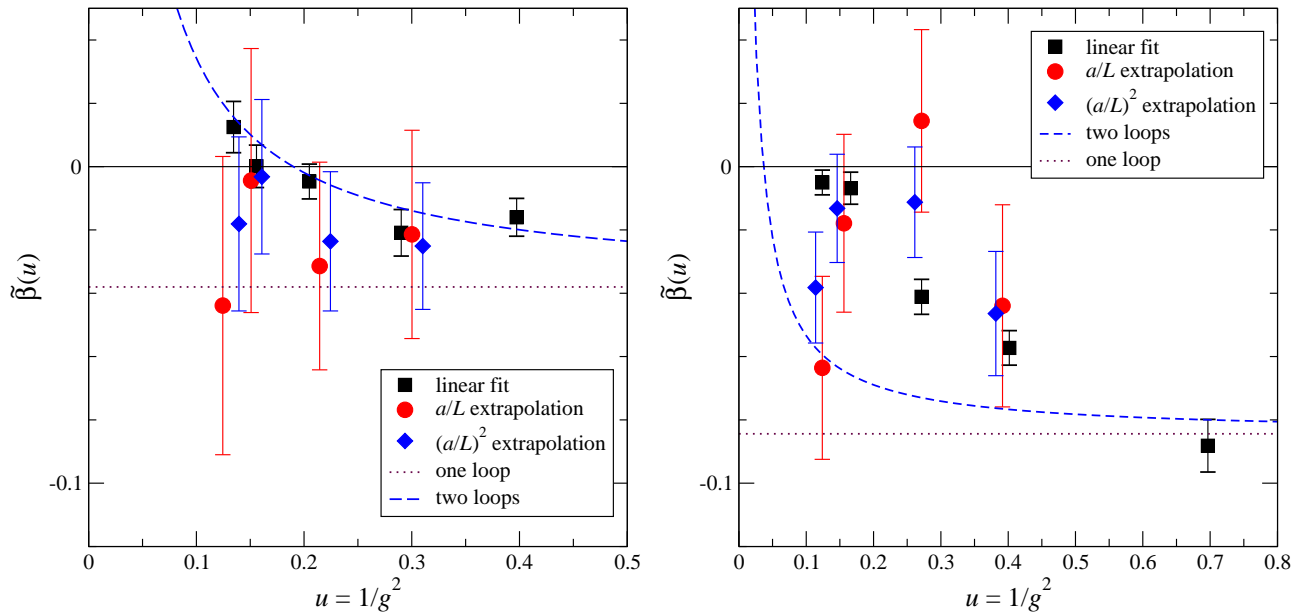

\begin{center}
\includegraphics[width=.47\textwidth,clip]{SU3adjbeta_scaled.eps}
\
\includegraphics[width=.47\textwidth,clip]{SU4as2beta.eps}
\end{center}
\caption{
Beta function $\tbeta(u)$ of the SU(3)/adjoint theory (left)
and the SU(4)/sextet theory (right), extrapolated to the continuum limit.
Black squares and curves are the same as in Fig.~\ref{fig:beta}.
The points for the extrapolations have been displaced slightly for clarity.
\label{fig:beta_cont}}
\end{figure*}

We plot the results of the continuum extrapolations for both the SU(3) and the SU(4) theories in Fig.~\ref{fig:beta_cont}.
%%%%%%%%%%%%%%%%%%%%%%%%%%%%%%%%%%%%%%%%%%%%%%%%%%%%%%%%%%%%%%%%%%%%%
%%%%%%%%%%%%%%%%%%%%%%%%%%%%%%%%%%%%%%%%%%%%%%%%%%%%%%%%%%%%%%%%%%%%%
Compared to Fig.~\ref{fig:beta}, the linear extrapolations increase the error bars by a factor of 5, the quadratic extrapolations by only a factor of 3.
The linear and quadratic extrapolations are mutually consistent for each data point;
one can consider them separately as distinct models, or take their error bars together as a combination of statistical with systematic errors.%
\footnote{For both theories we do not extrapolate the beta function at the weakest coupling; the absence of data for $L=16a$ leads to very large error in the extrapolation.}

We note that the simple linear fits (\ref{linfit}) typically give a large $\chi^2$ precisely because they neglect finite-lattice corrections.
The extrapolations  (\ref{linex}) and~(\ref{quadex}), on the other hand, are models aimed at
removing the discretization error and indeed they result in acceptable $\chi^2$.

In the SU(3) theory, we can no longer tell whether the beta function crosses zero, and indeed the very shape of the function is not well determined.
In the SU(4) theory, the extrapolations indicate a function that approaches zero and then veers off downwards.

%%%%%%%%%%%%%%%%%%%%%%%%%%%%%%%%%%%%%%%%%%%%%%%%%%%%%%%%%%%%%%%%%%%%%
\section{Mass anomalous dimension\label{sec:gamma}}
%%%%%%%%%%%%%%%%%%%%%%%%%%%%%%%%%%%%%%%%%%%%%%%%%%%%%%%%%%%%%%%%%%%%%

Following still the methods used in our previous work,
we calculate the mass anomalous dimension from the scaling with $L$ of the
pseudoscalar renormalization factor $Z_P$.
The latter comes from the ratio
\bee
Z_P = \frac {c \sqrt{f_1}}{f_P(L/2)}.
\label{eq:ZP}
\ee
$f_P$ is the propagator from a wall source at the $t=0$ boundary
to a point pseudoscalar operator at time $L/2$.
The normalization of the wall source is removed by the $\sqrt{f_1}$ factor,
which comes from a boundary-to-boundary correlator.
The constant $c$, which is an arbitrary normalization,
is $1/\sqrt{2}$ in our convention.

\begin{figure}
\begin{center}
\includegraphics[width=.95\columnwidth,clip]{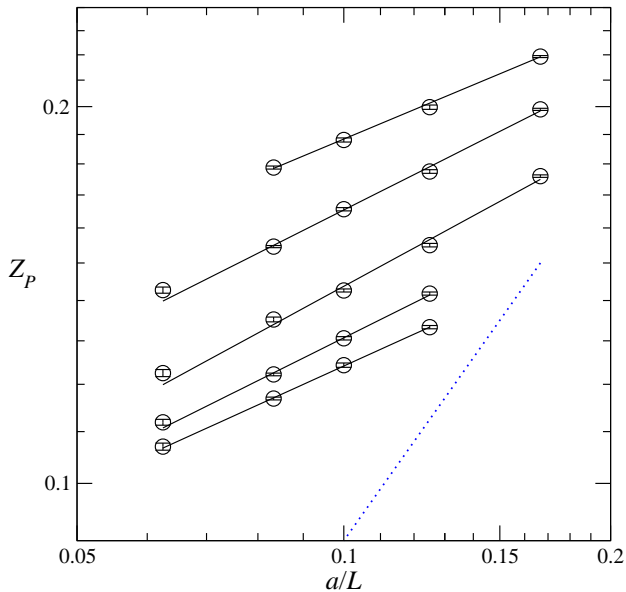}
\end{center}
\caption{The pseudoscalar renormalization constant
  $Z_P$ vs.~$L/a$ in the SU(3)/adjoint theory (Table~\ref{table:su3ZP}).
  Top to bottom: $\beta=5.0,$ 4.5, 4.1, 3.9, 3.8.
  The straight lines are linear fits to each set of points at given
  $\beta$; the slope gives $\gamma_m$.
  The hypothetical dotted line corresponds to $\gamma_m=1$.
\label{fig:su3ZP}}
\end{figure}

\begin{figure}
\begin{center}
\includegraphics[width=.95\columnwidth,clip]{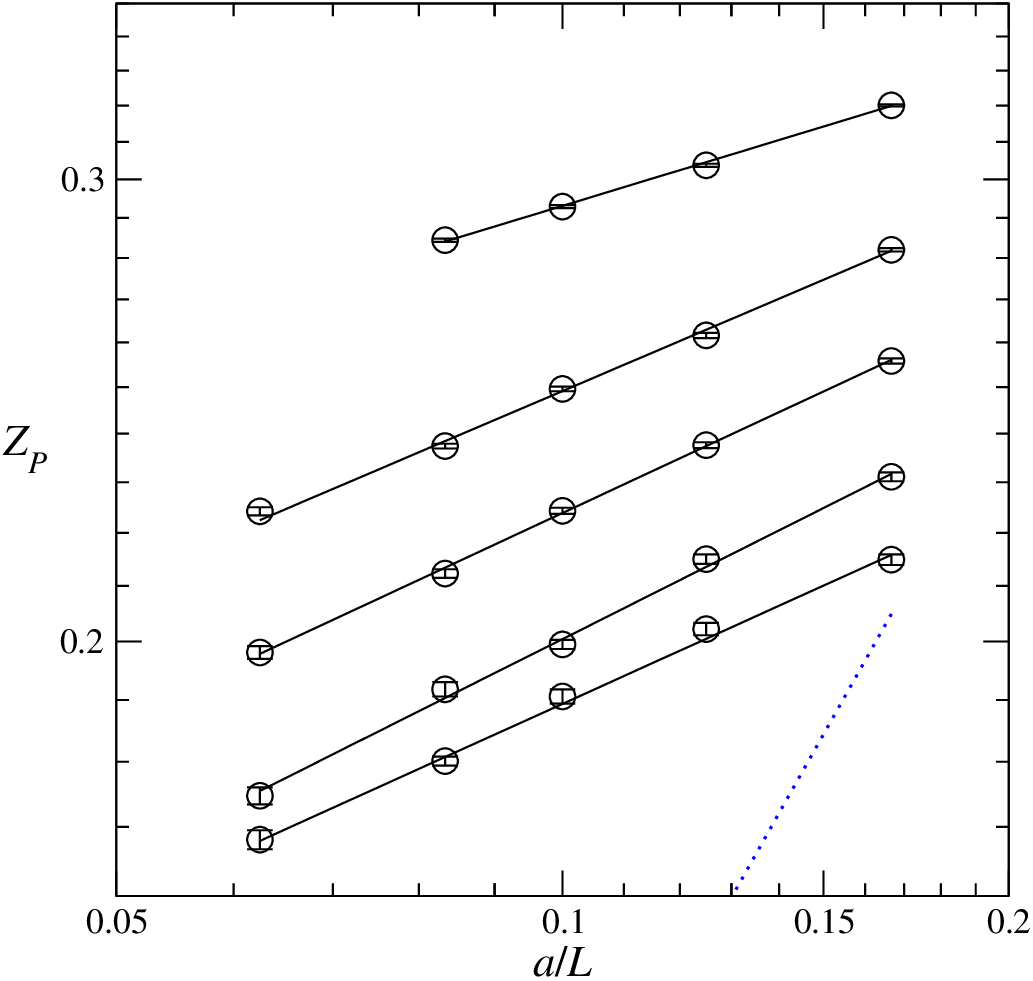}
\end{center}
\caption{
Same as Fig.~\ref{fig:su3ZP}, but for the SU(4)/sextet theory
  (Table~\ref{table:su4ZP}).
  Top to bottom: $\beta=10.0,$ 8.0, 7.0, 6.0, 5.5.
\label{fig:su4ZP}}
\end{figure}

We present in Tables~\ref{table:su3ZP} and~\ref{table:su4ZP} the values of
$Z_P$ we find in the SU(3) and SU(4) theories, respectively;
we plot them in Figs.~\ref{fig:su3ZP} and~\ref{fig:su4ZP}.

%%%%%%%%%%%%%%%%%%%%%%%%%%%%%%%%%%%%%%%%%%%%%%%%%%%%%%%%%%%%%%%%%%%%%%
\begin{table*}
\caption{Pseudoscalar renormalization constant $Z_P$ measured in the SU(3)/adjoint theory.}
\begin{center}
\begin{ruledtabular}
\begin{tabular}{ccllll}
$\beta$ &  \multicolumn{5}{c}{$Z_P$}\\
\cline{2-6}
    & $L=6a$     & $L=8a$     & $L=10a$    & $L=12a$    & $L=16a$  \\
\hline
3.8 &  \hfil--  & 0.1333(4) & 0.1243(5) & 0.1169(3) & 0.1070(7)%
\footnote{Average of three streams out of four; see Sec.~\ref{sec:action}.} \\
3.9 &  \hfil--  & 0.1418(4) & 0.1306(4) & 0.1222(3) & 0.1119(6) \\
4.1 & 0.1760(4) & 0.1550(5) & 0.1426(4) & 0.1352(6) & 0.1225(8) \\

4.5 & 0.1990(4) & 0.1775(6) & 0.1656(5) & 0.1546(3) & 0.1427(8) \\
5.0 & 0.2193(4) & 0.1998(8) & 0.1881(7) & 0.1788(5) &  \hfil--  \\
\end{tabular}
\end{ruledtabular}
\end{center}
\label{table:su3ZP}
\end{table*}
%%%%%%%%%%%%%%%%%%%%%%%%%%%%%%%%%%%%%%%%%%%%%%%%%%%%%%%%%%%%%%%%%%%%%
%%%%%%%%%%%%%%%%%%%%%%%%%%%%%%%%%%%%%%%%%%%%%%%%%%%%%%%%%%%%%%%%%%%%%
%%%%%%%%%%%%%%%%%%%%%%%%%%%%%%%%%%%%%%%%%%%%%%%%%%%%%%%%%%%%%%%%%%%%%
%%%%%%%%%%%%%%%%%%%%%%%%%%%%%%%%%%%%%%%%%%%%%%%%%%%%%%%%%%%%%%%%%%%%%%
\begin{table*}
\caption{Pseudoscalar renormalization constant $Z_P$ measured in the SU(4)/sextet theory.}
\begin{center}
\begin{ruledtabular}
\begin{tabular}{ccllll}
$\beta$ &  \multicolumn{5}{c}{$Z_P$}\\
\cline{2-6}
    & $L=6a$     & $L=8a$     & $L=10a$    & $L=12a$    & $L=16a$  \\
\hline
 5.5 & 0.2149(10) & 0.2022(11) & 0.1906(12) & 0.1801(7)  & 0.1681(14) \\
 6.0 & 0.2311(9)  & 0.2150(9)  & 0.1995(8)  & 0.1918(12) & 0.1747(13) \\
 7.0 & 0.2558(6)  & 0.2376(6)  & 0.2243(6)  & 0.2123(8)  & 0.1981(11) \\
 8.0 & 0.2820(4)  & 0.2616(6)  & 0.2496(5)  & 0.2374(5)  & 0.2242(8)  \\
10.0 & 0.3201(3)  & 0.3037(4)  & 0.2929(4)  & 0.2844(4)  &  \hfil--   \\
\end{tabular}
\end{ruledtabular}
\end{center}
\label{table:su4ZP}
\end{table*}
%%%%%%%%%%%%%%%%%%%%%%%%%%%%%%%%%%%%%%%%%%%%%%%%%%%%%%%%%%%%%%%%%%%%%
%%%%%%%%%%%%%%%%%%%%%%%%%%%%%%%%%%%%%%%%%%%%%%%%%%%%%%%%%%%%%%%%%%%%%
%%%%%%%%%%%%%%%%%%%%%%%%%%%%%%%%%%%%%%%%%%%%%%%%%%%%%%%%%%%%%%%%%%%%%

\begin{figure*}
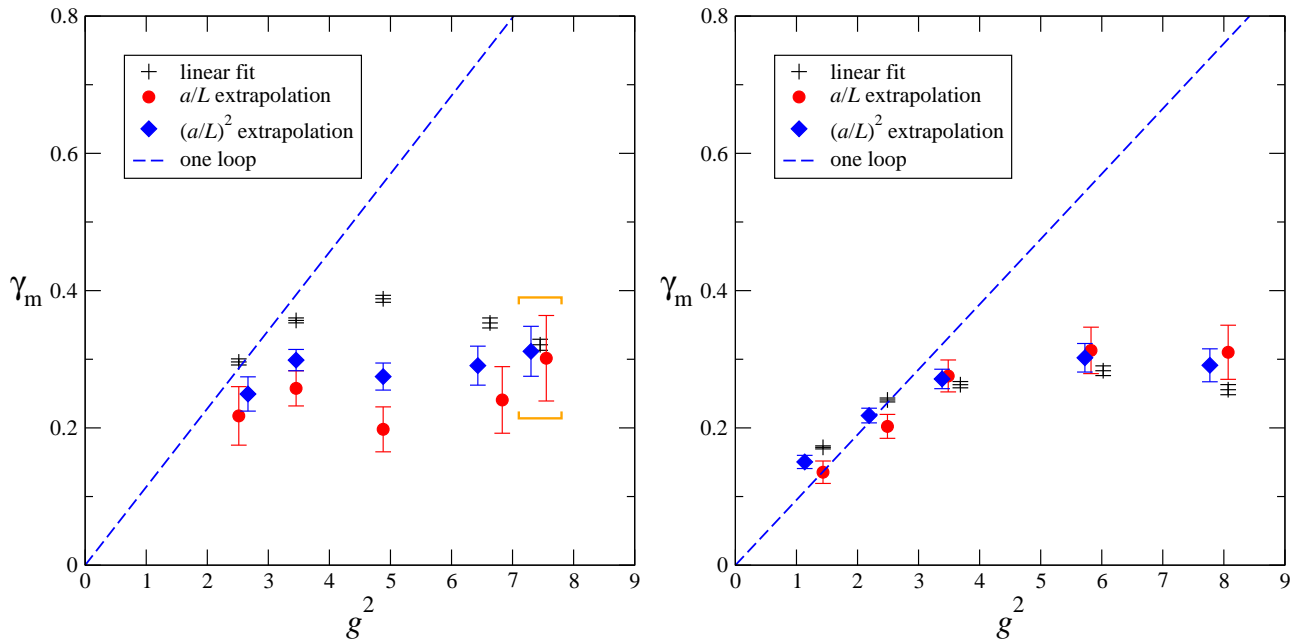

\begin{center}
\includegraphics[width=.47\textwidth,clip]{SU3adjgamma.eps}
\
\includegraphics[width=.47\textwidth,clip]{SU4as2gamma.eps}
\end{center}
\caption{
Mass anomalous dimension $\gamma_m(g^2)$ of the SU(3)/adjoint theory (left)
and the SU(4)/sextet theory (right), plotted as a function of $g^2(L=8a)$.
Shown are the simple linear fits and the linear and quadratic
extrapolations to the continuum limit.
The points for the extrapolations have been displaced slightly where necessary.
The brackets indicate the results of fits at $\beta=3.8$, which were obtained after dropping an outlying stream for $L=16a$; see Sec.~\ref{sec:action} and Table~\ref{table:su3ZP}.
(Restoring the dropped stream to the averages moves the bracketed points upwards slightly:
the simple linear fit by $1.4\sigma$, the linear extrapolation by $1.9\sigma$, and the quadratic extrapolation by $1.9\sigma$.)
\label{fig:gamma}}
\end{figure*}

As in the calculation of $\tbeta$, we begin with the simple fits, based on the slowness of the running of $1/g^2$~\cite{DeGrand:2011qd}.
Following the approximate scaling formula
\bee
Z_P(L)=Z_P(L_0)\left(\frac{L_0}L\right)^{\gamma_m} ,
\label{Zfp}
\ee
we fit the $Z_P$ data at each value of $\beta$ to
\bee
\log Z_P(L)=c_0+c_1 \log \frac {8a}L ,
\label{linfitZ}
\ee
giving the straight lines plotted in the figures; the slope $c_1$
gives an estimate of $\gamma_m$.
For an analysis of finite-volume effects, we drop successive volumes
starting from the smallest, giving the sequence of $c_1^{(n)}$ as above.
Again we extrapolate $c_1^{(n)}$ either linearly or quadratically to $a/L=0$.
All these results are plotted in Fig.~\ref{fig:gamma}.

In both theories, the simple linear fits produce values of $\gamma_m$ that depart from the one-loop line and level off.
In the SU(3) theory, the extrapolations drive the result downward.
Overall, we have a bound $\gamma_m\alt 0.4$.
In the SU(4) theory, the extrapolations are remarkably consistent with each other and with the original linear fit.
$\gamma_m$ again agrees well with the one-loop line in weak coupling, and then deviates downward to level off below 0.3 for the linear fits, stretching to 0.35 for the extrapolations.

The behavior of $\gamma_m$ in both theories is remarkably similar to our what we found in the three 2ISR theories: SU(2)/triplet, SU(3)/sextet, and SU(4)/decuplet.

%%%%%%%%%%%%%%%%%%%%%%%%%%%%%%%%%%%%%%%%%%%%%%%%%%%%%%%%%%%%%%%%%%%%%
%%%%%%%%%%%%%%%%%%%%%%%%%%%%%%%%%%%%%%%%%%%%%%%%%%%%%%%%%%%%%%%%%%%%%

%%%%%%%%%%%%%%%%%%%%%%%%%%%%%%%%%%%%%%%%%%%%%%%%%%%%%%%%%%%%%%%%%%%%%
\section{Conclusions\label{sec:concl}}
%%%%%%%%%%%%%%%%%%%%%%%%%%%%%%%%%%%%%%%%%%%%%%%%%%%%%%%%%%%%%%%%%%%%%

Our calculations reveal that  the beta functions associated with
the SF coupling in the two theories studied are small,
everywhere smaller than the one-loop values. In the SU(4) theory, the running is even slower than
what is expected in two loops. Our inability to disentangle possible lattice artifacts from
real running prevents a more definite statement.

In all cases we have studied, the two in this paper and the 2ISR theories in our previous work, the mass anomalous dimension
varies linearly with the SF gauge coupling when the coupling is small, and then levels off to a plateau at large gauge coupling. All the plateaus are at a level below 0.5.

Imagine now performing a lattice simulation for any of these systems at
any value of the bare lattice coupling in which the system is in the
same phase as at weak coupling. One will have access to physical scales
ranging from the lattice spacing $a$ to the system size $L$, where
(in the near future) $L/a$ will be smaller than about 100. The running coupling will
scarcely evolve over this change of scale. Whether or not the system is
actually at a fixed point, the slow evolution of the coupling implies
that lattice spectroscopy will display systematics of scaling, broken by
a nonzero fermion mass, by irrelevant operators, and by the effect of finite system volume.  Given
the size of the one-loop coefficient of the beta function, plus the
observation that the running coupling in these theories always runs more
slowly than one-loop expectations, this behavior is completely natural.

In all these lattice systems, typically one will encounter a confining
phase with broken chiral symmetry when the bare coupling exceeds a
certain value.  Whether or not this describes continuum physics
can in principle be decided by calculating the beta function
as we have attempted, and determining whether an IRFP is encountered
before chiral symmetry breaks.  In the two systems
studied in this paper, we were unable to resolve this question.

The great advantage of lattice QCD is that the bare coupling can be
adjusted such that perturbation theory is valid at the lattice scale $a$,
while at the same time the volume is big enough to accommodate even
the lightest of the hadrons.
The same is not true for nearly conformal theories.  Whether the infrared
physics is conformal or not, in order to probe it the bare coupling
must be strong.
One consequence is that the Symanzik effective action, defined around the gaussian fixed point, offers no guidance to the scaling dimensions of irrelevant operators.
In a nearly conformal theory, where finite-lattice corrections are essentially functions of $a/L$, this also leaves us ignorant regarding the behavior of finite-volume corrections.
Our extrapolations to infinite volume are then only models.

A comparison to the SF analysis of ordinary QCD (with small $N_f$ and triplet quarks) invites the question of why it is so difficult to produce good quality results for borderline-conformal theories. We believe that the
answer lies in the fact that what is interesting is not the absolute uncertainty $\Delta \tbeta$ in
the value of the beta function $\tbeta(g^2)$; rather, it is the relative error, for which we take the ratio of the uncertainty to the one-loop constant value.
The latter is proportional to the lowest order coefficient $b_1$.
In QCD with three flavors, $b_1=9$; for the near-conformal theories, Table~\ref{table:bfun} lists values that are a good deal smaller.
The QCD beta function is also increased by a positive $b_2$, whereas $b_2<0$ is a necessary feature of the borderline theories.
Indeed, Table~\ref{table:bfun} shows that the SU(3)/adjoint theory studied here is a particularly difficult case to begin with.

The uncertainty $\Delta \tbeta$ scales with the ensemble
size as $1/\sqrt{N}$, where $N$ is the number of uncorrelated measurements.
The observable giving the SF coupling is essentially a surface quantity.
It also includes data generated by a noisy estimator.
Thus it has large inherent fluctuations as well as long-time autocorrelations underlying these fluctuations.
While this is true for all theories, we found the SU(3)/adjoint model to be particularly intractable.
New methods of computing running couplings will compete successfully with the SF if they can overcome these problems.

%%%%%%%%%%%%%%%%%%%%%%%%%%%%%%%%%%%%%%%%%%%%%%%%%%%%%%%%%%%%%%%%%%%%%
\begin{acknowledgments}
We thank Steven Gottlieb, Don Holmgren, and James Simone for assistance.
Y.~S. and B.~S. thank the Galileo Galilei Institute in Florence for its hospitality; Y.~S. thanks the University of Colorado similarly.
This work was supported in part by the Israel Science Foundation
under grant no.~423/09 and by the U.~S. Department of Energy under grant DE-FG02-04ER41290.

Our computational work used the Extreme Science and Engineering Discovery Environment (XSEDE), which is supported by National Science Foundation grant number OCI-1053575.
The computations were carried out at
(1) the University of Texas and
(2) the National Institute for Computational Sciences (NICS) at the University of Tennessee,
under XSEDE project number TG-PHY090023.

We also enjoyed a substantial grant of computer time from
the QCD Lattice Group at Fermilab,
whose facilities are funded by the Office of Science of the U.~S. Department of Energy and allocated via the USQCD collaboration.
Additional computation time was granted by: (1) the LinkSCEEM-2 project, funded by the European Commission under the 7th Framework Programme through Capacities Research Infrastructure, INFRA-2010-1.2.3 Virtual Research Communities, Combination of Collaborative Project and Coordination and Support Actions (CP-CSA) under grant agreement no RI-261600;
and (2) the High-Performance Computing Infrastructure for South East Europe's Research Communities (HP-SEE), a project co-funded by the European Commission (under contract number 261499) through the Seventh Framework Programme.

Our computer code is based on the publicly available package of the
 MILC collaboration~\cite{MILC}.
The code for hypercubic smearing was adapted from a program written by A.~Hasenfratz,
R.~Hoffmann and S.~Schaefer~\cite{Hasenfratz:2007rf}.

\end{acknowledgments}
%%%%%%%%%%%%%%%%%%%%%%%%%%%%%%%%%%%%%%%%%%%%%%%%%%%%%%%%%%%%%%%%%%%%%
\appendix* \section{}
\subsection{Choosing the value of $\beta'$}
\begin{figure}
\begin{center}
\includegraphics[width=.95\columnwidth,clip]{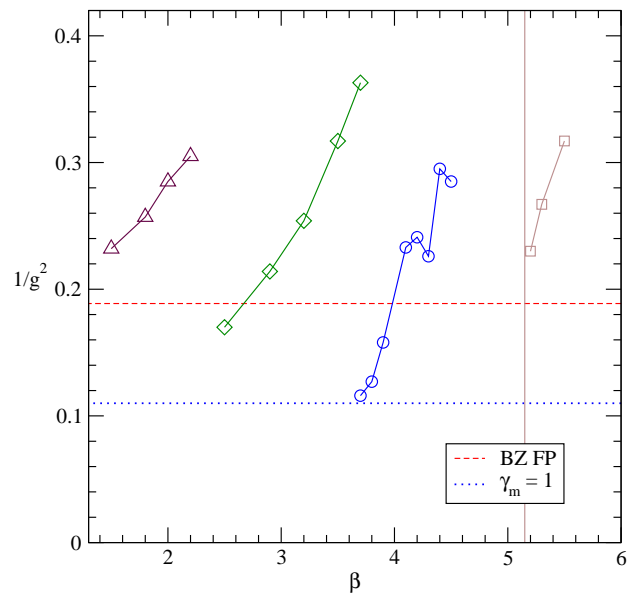}
\end{center}
\caption{
Inverse SF coupling versus gauge coupling $\beta$ for several choices
 of $\beta'$ in the SU(3)/adjoint theory, measured with short runs on a $6^4$ lattice.
The connected data sets are for (right to left) $\beta'=-0.5$, 0, 0.5, 1.0.
For  $\beta'=-0.5$, the vertical line marks the
appearance of the first-order transition that makes $\kappa_c$ disappear for smaller $\beta$.
The horizontal dashed line near the bottom of the graph marks the
 location of the Banks--Zaks (two-loop) fixed point.
The horizontal dotted line marks where the one-loop $\gamma_m(g^2)$ is equal to unity.
Statistical error bars range from $\pm0.01$ to $\pm0.02$.
\label{fig:1g2SU3}}
\end{figure}

\begin{figure}
\begin{center}
\includegraphics[width=.95\columnwidth,clip]{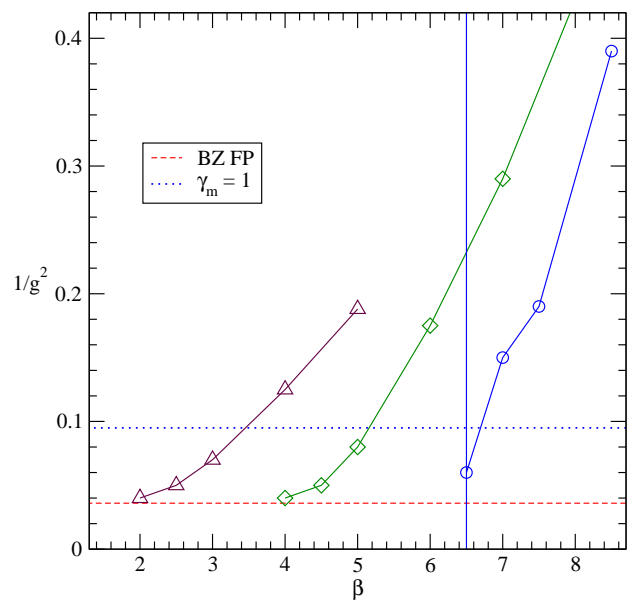}
\end{center}
\caption{
As in Fig.~\ref{fig:1g2SU3}, but
in the SU(4)/sextet theory.
The connected data sets are for (right to left) $\beta'=0$, 0.5, 1.0.
The vertical line marks the
appearance of the first-order transition for $\beta'=0$.
\label{fig:1g2SU4}}
\end{figure}

%%%%%%%%%%%%%%%%%%%%%%%%%%%%%%%%%%%%%%%%%%%%%%%%%%%%%%%%%%%%%%%%%%%%%
To find an optimal value of $\beta'$ for each theory, we do a series of short runs on small ($L=6a$) lattices to determine $\kappa_c(\beta)$ and then to measure the SF coupling $g^2$ along the $\kappa_c$ curve.
For fixed $\beta'$, $g^2$ grows as $\beta$ is decreased (see Figs.~\ref{fig:1g2SU3} and~\ref{fig:1g2SU4}).
Our aim is to reach the largest $g^2$ possible.
This is limited by two effects that appear as $\beta$ is decreased:
Either one reaches the point $\beta_1$ where the $\kappa_c(\beta)$ curve hits the first-order transition seen in Fig.~\ref{fig:sketch}, or the poor acceptance due to the increasing disorder makes simulation impractical.
The encounter with the phase transition is indicated by vertical lines drawn for one value of $\beta'$ in each of Figs.~\ref{fig:1g2SU3} and~\ref{fig:1g2SU4}; for other values of $\beta'$, we were stopped by acceptance so poor on these small lattices that simulation on larger lattices would have been impossible.
%%%%%%%%%%%%%%%%%%%%%%%%%%%%%%%%%%%%%%%%%%%%%%%%%%%%%%%%%%%%%%%%%%%%%
%%%%%%%%%%%%%%%%%%%%%%%%%%%%%%%%%%%%%%%%%%%%%%%%%%%%%%%%%%%%%%%%%%%%%
%%%%%%%%%%%%%%%%%%%%%%%%%%%%%%%%%%%%%%%%%%%%%%%%%%%%%%%%%%%%%%%%%%%%%
%%%%%%%%%%%%%%%%%%%%%%%%%%%%%%%%%%%%%%%%%%%%%%%%%%%%%%%%%%%%%%%%%%%%%

For the SU(3)/adjoint theory we chose $\beta'=0$ since it appeared to offer the largest range in $g^2$.
As it turned out, longer runs on $6^4$ lattices showed that the three points plotted for the strongest couplings ($\beta=3.7$, 3.8,~3.9) in Fig.~\ref{fig:1g2SU3} represented metastable states, i.e., lying on the wrong side of the phase transition.
We were nonetheless able to run at $\beta=3.8$ and~3.9 on larger lattices, as described in Sec.~\ref{sec:action}.
In the SU(4)/sextet theory no such issue arose; Fig.~\ref{fig:1g2SU4} shows why we chose $\beta'=0.5$ for this theory.

Figures \ref{fig:1g2SU3} and~\ref{fig:1g2SU4} may be compared to Fig.~1 in our paper on the SU(4)/decuplet theory~\cite{DeGrand:2012qa}.
One may find there a demonstration of universality in weak coupling as $\beta'$ is varied.

%%%%%%%%%%%%%%%%%%%%%%%%%%%%%%%%%%%%%%%%%%%%%%%%%%%%%%%%%%%%%%%%%%%%%
\subsection{Continuum extrapolations}
%%%%%%%%%%%%%%%%%%%%%%%%%%%%%%%%%%%%%%%%%%%%%%%%%%%%%%%%%%%%%%%%%%%%%
%%%%%%%%%%%%%%%%%%%%%%%%%%%%%%%%%%%%%%%%%%%%%%%%%%%%%%%%%%%%%%%%%%%%%
\begin{figure*}
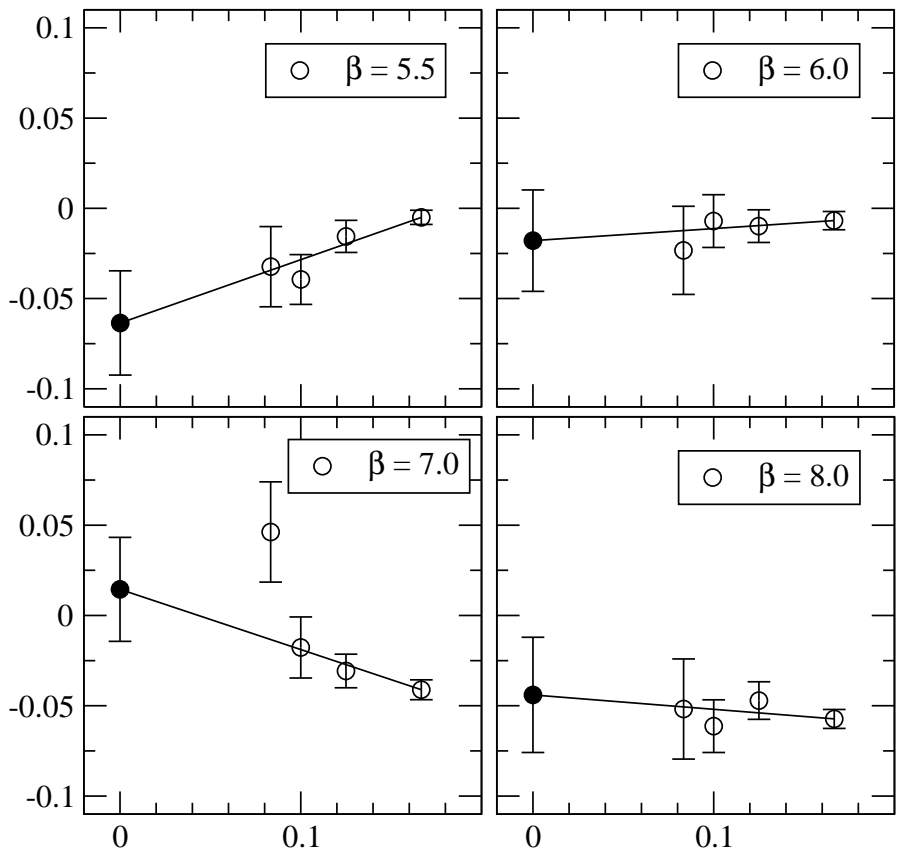

%\begin{center}
\includegraphics[height=.3\textwidth,clip]{dfl55.eps}
\includegraphics[height=.3\textwidth,clip]{dfl60.eps}\hfill\\
\includegraphics[height=.325\textwidth,clip]{dfl70.eps}
\includegraphics[height=.325\textwidth,clip]{dfl80.eps}
%\end{center}
\caption{
Successive fits $c_1^{(n)}$ giving the beta function $\tbeta$ of the SU(4)/sextet theory, as a function of $a/L_n$, where $L_n$ is the smallest lattice size used in the fit.  The linear extrapolations to $a/L=0$ are shown.
\label{fig:effbeta_l}}
\end{figure*}
%%%%%%%%%%%%%%%%%%%%%%%%%%%%%%%%%%%%%%%%%%%%%%%%%%%%%%%%%%%%%%%%%%%%%
%%%%%%%%%%%%%%%%%%%%%%%%%%%%%%%%%%%%%%%%%%%%%%%%%%%%%%%%%%%%%%%%%%%%%
\begin{figure*}
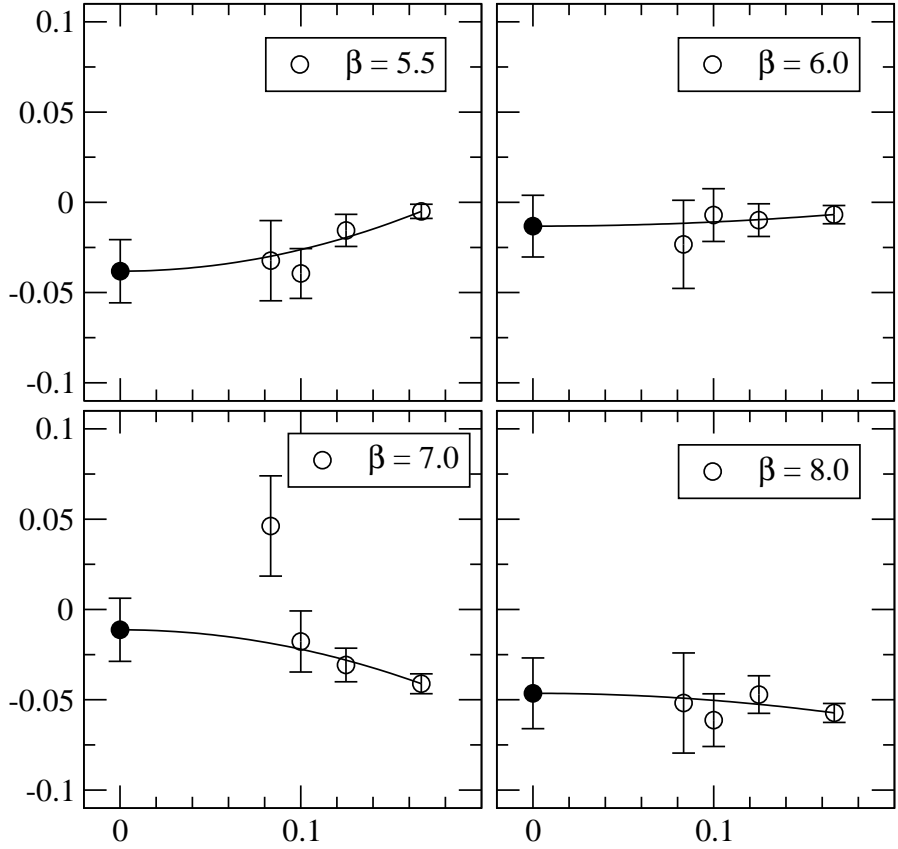

%\begin{center}
\includegraphics[height=.3\textwidth,clip]{dfq55.eps}
\includegraphics[height=.3\textwidth,clip]{dfq60.eps}\hfill\\
\includegraphics[height=.325\textwidth,clip]{dfq70.eps}
\includegraphics[height=.325\textwidth,clip]{dfq80.eps}
%\end{center}
\caption{
Same as Fig.~\ref{fig:effbeta_l}, but showing the quadratic extrapolations to $a/L=0$.
\label{fig:effbeta_q}}
\end{figure*}
%%%%%%%%%%%%%%%%%%%%%%%%%%%%%%%%%%%%%%%%%%%%%%%%%%%%%%%%%%%%%%%%%%%%%
To illustrate our method of continuum extrapolation, we show the values of $c_1^{(n)}$ for the beta function in the SU(4) theory in Figs.~\ref{fig:effbeta_l} and~\ref{fig:effbeta_q}, along with their linear and quadratic extrapolations to $a/L=0$.
The figures show the origin of the error bars in the extrapolated values at $L\to\infty$.
The quadratic extrapolations result in smaller error bars because they have a longer lever arm between the smallest and largest lattices.

%%%%%%%%%%%%%%%%%%%%%%%%%%%%%%%%%%%%%%%%%%%%%%%%%%%%%%%%%%%%%%%%%%%%%

%%%%%%%%%%%%%%%%%%%%%%%%%%%%%%%%%%%%%%%%%%%%%%%%%%%%%%%%%

\begin{thebibliography}{99}  
%%%%%%%%%%%%%%%%%%%%%%%%%%%%%%%%%%%%%%%%%%%%%%%%%%%%%%%%%%%%%%%%%%%%%

%%%%%%%%%%%%%%%%%%%%%%%%%%%%%%%
% introduction refs
%%%%%%%%%%%%%%%%%%%%%%%%%%%%%%%

%%%%%%%%% Reviews %%%%%%%%%%%%%

%\cite{Neil:2012cb}
\bibitem{Neil:2012cb}
  E.~T.~Neil,
  ``Exploring Models for New Physics on the Lattice,''
  PoS LATTICE {\bf 2011}, 009 (2011)
  [arXiv:1205.4706 [hep-lat]].
  %%CITATION = ARXIV:1205.4706;%%

%\cite{Giedt:2012it}
\bibitem{Giedt:2012it}
  J.~Giedt,
  ``Lattice gauge theory and physics beyond the standard model,''
  PoS LATTICE {\bf 2012}, 006 (2012).
  %%CITATION = POSCI,LATTICE2012,006;%%

%%%%%%%%% walking %%%%%%%%%%%%%

%\cite{Holdom:1981rm}
\bibitem{Holdom:1981rm}
  B.~Holdom,
  \ttl{Raising The Sideways Scale,}
  Phys.\ Rev.\  D {\bf 24}, 1441 (1981).
  %%CITATION = PHRVA,D24,1441;%%

%\cite{Yamawaki:1985zg}
\bibitem{Yamawaki:1985zg}
  K.~Yamawaki, M.~Bando and K.~i.~Matumoto,
  \ttl{Scale Invariant Technicolor Model And A Technidilaton,}
  Phys.\ Rev.\ Lett.\  {\bf 56}, 1335 (1986).
  %%CITATION = PRLTA,56,1335;%%

%\cite{Hill:2002ap}
\bibitem{Hill:2002ap}
  C.~T.~Hill and E.~H.~Simmons,
  \ttl{Strong dynamics and electroweak symmetry breaking,}
  Phys.\ Rept.\  {\bf 381}, 235 (2003)
  [Erratum-ibid.\  {\bf 390}, 553 (2004)]
  [arXiv:hep-ph/0203079];
%%CITATION = PRPLC,381,235;%%

%%%%%%%%% beta fns %%%%%%%%%%%%%

\bibitem{Sannino:2004qp}
F.~Sannino and K.~Tuominen,
  \ttl{Orientifold theory dynamics and symmetry breaking,}
  Phys.\ Rev.\  D {\bf 71}, 051901 (2005)
  [arXiv:hep-ph/0405209].
  %%CITATION = PHRVA,D71,051901;%%

%\cite{Hong:2004td}
\bibitem{Hong:2004td}
  D.~K.~Hong, S.~D.~H.~Hsu and F.~Sannino,
  \ttl{Composite Higgs from higher representations,}
  Phys.\ Lett.\  B {\bf 597}, 89 (2004)
  [arXiv:hep-ph/0406200].
%%CITATION = PHLTA,B597,89;%%

%\cite{Dietrich:2006cm}
\bibitem{Dietrich:2006cm}
  D.~D.~Dietrich and F.~Sannino,
  ``Conformal window of SU($N$) gauge theories with fermions in higher dimensional representations,''
  Phys.\ Rev.\ D {\bf 75}, 085018 (2007)
  [hep-ph/0611341].
  %%CITATION = HEP-PH/0611341;%%

\bibitem{Caswell:1974gg}
  W.~E.~Caswell,
  \ttl{Asymptotic behavior of nonabelian gauge theories to two loop order,}
  Phys.\ Rev.\ Lett.\  {\bf 33}, 244 (1974).
  %%CITATION = PRLTA,33,244;%%

\bibitem{Banks:1981nn}
  T.~Banks and A.~Zaks,
  \ttl{On the phase structure of vector-like gauge theories with massless fermions,}
  Nucl.\ Phys.\  B {\bf 196}, 189 (1982).
  %%CITATION = NUPHA,B196,189;%%

%\cite{Appelquist:1988yc}
\bibitem{Appelquist:1988yc}
  T.~Appelquist, K.~D.~Lane and U.~Mahanta,
  ``On The Ladder Approximation For Spontaneous Chiral Symmetry Breaking,''
  Phys.\ Rev.\ Lett.\  {\bf 61}, 1553 (1988).
  %%CITATION = PRLTA,61,1553;%%

\bibitem{Cohen:1988sq}
  A.~G.~Cohen, and H.~Georgi,
  ``Walking Beyond The Rainbow,''
  Nucl.\ Phys.\  B {\bf 314}, 7 (1989).
  %%CITATION = NUPHA,B314,7;%%

%%%%%%%%% SF %%%%%%%%%%%%%

\bibitem{Luscher:1992an}
  M.~L\"uscher, R.~Narayanan, P.~Weisz and U.~Wolff,
  \ttl{The Schr\"odinger functional: A Renormalizable probe for non-Abelian gauge theories,}
  Nucl.\ Phys.\  B {\bf 384}, 168 (1992)
  [arXiv:hep-lat/9207009].
  %%CITATION = NUPHA,B384,168;%%

\bibitem{Luscher:1993gh}
  M.~L\"uscher, R.~Sommer, P.~Weisz and U.~Wolff,
  \ttl{A precise determination of the running coupling in the SU(3) Yang-Mills theory,}
  Nucl.\ Phys.\  B {\bf 413}, 481 (1994)
  [arXiv:hep-lat/9309005].
  %%CITATION = NUPHA,B413,481;%%

\bibitem{Sint:1993un}
  S.~Sint,
  ``On the Schr\"odinger functional in QCD,''
  Nucl.\ Phys.\  B {\bf 421}, 135 (1994)
  [arXiv:hep-lat/9312079];
  %%CITATION = NUPHA,B421,135;%%
  ``One Loop Renormalization Of The QCD Schr\"odinger Functional,''
  {\bf 451}, 416 (1995)
  [arXiv:hep-lat/9504005].
  %%CITATION = NUPHA,B451,416;%%

\bibitem{Sint:1995ch}
  S.~Sint and R.~Sommer,
  \ttl{The running coupling from the QCD Schr\"odinger functional: A one loop analysis,}
  Nucl.\ Phys.\  B {\bf 465}, 71 (1996)
  [arXiv:hep-lat/9508012].
  %%CITATION = NUPHA,B465,71;%%

\bibitem{Jansen:1998mx}
  K.~Jansen and R.~Sommer  [ALPHA collaboration],
  \ttl{O($\alpha$) improvement of lattice QCD with two flavors of Wilson quarks,}
  Nucl.\ Phys.\  B {\bf 530}, 185 (1998)
  [Erratum-{\em ibid.}\  B {\bf 643}, 517 (2002)]
  [arXiv:hep-lat/9803017].
  %%CITATION = NUPHA,B530,185;%%

\bibitem{DellaMorte:2004bc}
  M.~Della Morte {\em et al.} [ALPHA Collaboration],
  \ttl{Computation of the strong coupling in QCD with two dynamical flavours,}
  Nucl.\ Phys.\  B {\bf 713}, 378 (2005)
  [arXiv:hep-lat/0411025].
  %%CITATION = NUPHA,B713,378;%%

%%%%%%%%% us %%%%%%%%%%%%%%%%%%

%\cite{DeGrand:2011qd}
\bibitem{DeGrand:2011qd}
  T.~DeGrand, Y.~Shamir, B.~Svetitsky,
  \ttl{Infrared fixed point in SU(2) gauge theory with adjoint fermions,}
  Phys.\ Rev.\  {\bf D83}, 074507 (2011).
  [arXiv:1102.2843 [hep-lat]].
%%CITATION = ARXIV:1102.2843;%%

%\cite{Shamir:2008pb}
\bibitem{Shamir:2008pb}
  Y.~Shamir, B.~Svetitsky and T.~DeGrand,
  \ttl{Zero of the discrete beta function in SU(3) lattice gauge theory with color sextet fermions,}
  Phys.\ Rev.\  D {\bf 78}, 031502 (2008)
  [arXiv:0803.1707 [hep-lat]].
  %%CITATION = PHRVA,D78,031502;%%

%\cite{DeGrand:2010na}
\bibitem{DeGrand:2010na}
  T.~DeGrand, Y.~Shamir, B.~Svetitsky,
  \ttl{Running coupling and mass anomalous dimension of SU(3) gauge theory with two flavors of symmetric-representation fermions,}
  Phys.\ Rev.\  {\bf D82}, 054503 (2010).
  [arXiv:1006.0707 [hep-lat]].
%%CITATION = ARXIV:1006.0707;%%

%\cite{DeGrand:2012yq}
\bibitem{DeGrand:2012yq}
  T.~DeGrand, Y.~Shamir and B.~Svetitsky,
  ``Mass anomalous dimension in sextet QCD,''
    Phys.\ Rev.\ D {\bf 87}, 074507 (2013)
  [arXiv:1201.0935 [hep-lat]].
  %%CITATION = ARXIV:1201.0935;%%

%\cite{DeGrand:2012qa}
\bibitem{DeGrand:2012qa}
  T.~DeGrand, Y.~Shamir and B.~Svetitsky,
  ``SU(4) lattice gauge theory with decuplet fermions: Schr\"odinger functional analysis,''
    Phys.\ Rev.\ D {\bf 85}, 074506 (2012)
  [arXiv:1202.2675 [hep-lat]].
%%CITATION = ARXIV:1202.2675;%%

%%%%%%%%% other BSM SF %%%%%%%%%%%%%%%%%%

%\cite{Appelquist:2007hu}
\bibitem{Appelquist:2007hu}
  T.~Appelquist, G.~T.~Fleming and E.~T.~Neil,
  \ttl{Lattice study of the conformal window in QCD-like theories,}
  Phys.\ Rev.\ Lett.\  {\bf 100}, 171607 (2008)
  [Erratum-ibid.\  {\bf 102}, 149902 (2009)]
  [arXiv:0712.0609 [hep-ph]].
  %%CITATION = ARXIV:0712.0609;%%

%\cite{Appelquist:2009ty}
\bibitem{Appelquist:2009ty}
  T.~Appelquist, G.~T.~Fleming and E.~T.~Neil,
  \ttl{Lattice Study of Conformal Behavior in SU(3) Yang-Mills Theories,}
  Phys.\ Rev.\ D {\bf 79}, 076010 (2009)
  [arXiv:0901.3766 [hep-ph]].
  %%CITATION = ARXIV:0901.3766;%%

%\cite{Hietanen:2009az}
\bibitem{Hietanen:2009az}
  A.~J.~Hietanen, K.~Rummukainen and K.~Tuominen,
  \ttl{Evolution of the coupling constant in SU(2) lattice gauge theory with two adjoint fermions,}
  Phys.\ Rev.\ D {\bf 80}, 094504 (2009)
  [arXiv:0904.0864 [hep-lat]].
  %%CITATION = ARXIV:0904.0864;%%

%\cite{Bursa:2009we}
\bibitem{Bursa:2009we}
  F.~Bursa, L.~Del Debbio, L.~Keegan, C.~Pica and T.~Pickup,
  \ttl{Mass anomalous dimension in SU(2) with two adjoint fermions,}
  Phys.\ Rev.\  D {\bf 81}, 014505 (2010)
  [arXiv:0910.4535 [hep-ph]].
  %%CITATION = PHRVA,D81,014505;%%

%\cite{Bursa:2010xn}
\bibitem{Bursa:2010xn}
  F.~Bursa, L.~Del Debbio, L.~Keegan, C.~Pica and T.~Pickup,
  \ttl{Mass anomalous dimension in SU(2) with six fundamental fermions,}
  Phys.\ Lett.\ B {\bf 696}, 374 (2011)
  [arXiv:1007.3067 [hep-ph]].
  %%CITATION = ARXIV:1007.3067;%%

%\cite{Hayakawa:2010yn}
\bibitem{Hayakawa:2010yn}
  M.~Hayakawa, K.~-I.~Ishikawa, Y.~Osaki, S.~Takeda, S.~Uno and N.~Yamada,
  \ttl{Running coupling constant of ten-flavor QCD with the Schr\"odinger functional method,}
  Phys.\ Rev.\ D {\bf 83}, 074509 (2011)
  [arXiv:1011.2577 [hep-lat]].
  %%CITATION = ARXIV:1011.2577;%%

%\cite{Karavirta:2011zg}
\bibitem{Karavirta:2011zg}
  T.~Karavirta, J.~Rantaharju, K.~Rummukainen and K.~Tuominen,
  \ttl{Determining the conformal window: SU(2) gauge theory with $N_f = 4$, 6 and 10 fermion flavours,}
  JHEP {\bf 1205}, 003 (2012)
  [arXiv:1111.4104 [hep-lat]].
  %%CITATION = ARXIV:1111.4104;%%

%\cite{Hayakawa:2012gf}
\bibitem{Hayakawa:2012gf} 
  M.~Hayakawa, K.~-I.~Ishikawa, Y.~Osaki, S.~Takeda and N.~Yamada,
  ``Lattice study on two-color QCD with six flavors of dynamical quarks,''
  PoS LATTICE {\bf 2012}, 040 (2012)
  [arXiv:1210.4985 [hep-lat]].
  %%CITATION = ARXIV:1210.4985;%%

%\cite{Voronov:2012qx}
\bibitem{Voronov:2012qx} 
  G.~Voronov,
  ``Two-Color Schr\"odinger Functional with Six Flavors of Stout-Smeared Wilson Fermions,''
  PoS LATTICE {\bf 2012}, 039 (2012)
  [arXiv:1212.1376 [hep-lat]].
  %%CITATION = ARXIV:1212.1376;%%

%\cite{Rantaharju:2013gz}
\bibitem{Rantaharju:2013gz} 
  J.~Rantaharju, K.~Rummukainen and K.~Tuominen,
  ``Running coupling in SU(2) with adjoint fermions,''
  arXiv:1301.2373 [hep-lat].
  %%CITATION = ARXIV:1301.2373;%%

%%%%%%%%% ZP %%%%%%%%%%%%%%%%%%

%\cite{Sint:1998iq}
\bibitem{Sint:1998iq}
  S.~Sint and P.~Weisz  [ALPHA collaboration],
  \ttl{The running quark mass in the SF scheme and its two-loop anomalous dimension,}
  Nucl.\ Phys.\  B {\bf 545}, 529 (1999)
  [arXiv:hep-lat/9808013].
  %%CITATION = NUPHA,B545,529;%%

%\cite{Capitani:1998mq}
\bibitem{Capitani:1998mq}
  S.~Capitani, M.~L\"uscher, R.~Sommer and H.~Wittig  [ALPHA Collaboration],
  \ttl{Non-perturbative quark mass renormalization in quenched lattice QCD,}
  Nucl.\ Phys.\  B {\bf 544}, 669 (1999)
  [arXiv:hep-lat/9810063].
  %%CITATION = NUPHA,B544,669;%%

%\cite{DellaMorte:2005kg}
\bibitem{DellaMorte:2005kg}
  M.~Della Morte {\em et al.} %R.~Hoffmann, F.~Knechtli, J.~Rolf, R.~Sommer, I.~Wetzorke and U.~Wolff
  [ALPHA Collaboration],
  \ttl{Non-perturbative quark mass renormalization in two-flavor QCD,}
  Nucl.\ Phys.\  B {\bf 729}, 117 (2005)
  [arXiv:hep-lat/0507035].
  %%CITATION = NUPHA,B729,117;%%
  
%%%%%%%%% SU(3)/adj %%%%%%%%%%%%%%%%%%

%\cite{Kogut:1984sb}
\bibitem{Kogut:1984sb}
  J.~B.~Kogut, J.~Shigemitsu and D.~K.~Sinclair,
  ``Chiral Symmetry Breaking With Octet And Sextet Quarks,''
  Phys.\ Lett.\ B {\bf 145}, 239 (1984).
  %%CITATION = PHLTA,B145,239;%%

%\cite{Gerstenmayer:1989qw}
\bibitem{Gerstenmayer:1989qw}
  E.~Gerstenmayer, M.~Faber, W.~Feilmair, H.~Markum and M.~M\"uller,
  ``Quark Confinement With Dynamical Fermions In Higher Representations,''
  Phys.\ Lett.\ B {\bf 231}, 453 (1989).
  %%CITATION = PHLTA,B231,453;%%

\bibitem{Karsch:1998qj}
  F.~Karsch and M.~L\"utgemeier,
  ``Deconfinement and chiral symmetry restoration in an SU(3) gauge theory with adjoint fermions,''
  Nucl.\ Phys.\ B {\bf 550}, 449 (1999)
  [hep-lat/9812023].
  %%CITATION = HEP-LAT/9812023;%%

%\cite{Peskin:1980gc}
\bibitem{Peskin:1980gc}
  M.~E.~Peskin,
  ``The Alignment of the Vacuum in Theories of Technicolor,''
  Nucl.\ Phys.\ B {\bf 175}, 197 (1980).
  %%CITATION = NUPHA,B175,197;%%

%\cite{Basile:2004wa}
\bibitem{Basile:2004wa}
  F.~Basile, A.~Pelissetto and E.~Vicari,
  ``The Finite-temperature chiral transition in QCD with adjoint fermions,''
  JHEP {\bf 0502}, 044 (2005)
  [hep-th/0412026].
  %%CITATION = HEP-TH/0412026;%%

%\cite{Engels:2005te}
\bibitem{Engels:2005te}
  J.~Engels, S.~Holtmann and T.~Schulze,
  ``Scaling and Goldstone effects in a QCD with two flavors of adjoint quarks,''
  Nucl.\ Phys.\ B {\bf 724}, 357 (2005)
  [hep-lat/0505008].
  %%CITATION = HEP-LAT/0505008;%%

%\cite{Cossu:2008wh}
\bibitem{Cossu:2008wh}
  G.~Cossu, M.~D'Elia, A.~Di Giacomo, G.~Lacagnina and C.~Pica,
  ``Monopole condensation in two-flavor adjoint QCD,''
  Phys.\ Rev.\ D {\bf 77}, 074506 (2008)
  [arXiv:0802.1795 [hep-lat]].
  %%CITATION = ARXIV:0802.1795;%%

%\cite{Unsal:2007jx}
\bibitem{Unsal:2007jx}
  M.~Unsal,
  ``Magnetic bion condensation: A new mechanism of confinement and mass gap in four dimensions,''
  Phys.\ Rev.\ D {\bf 80}, 065001 (2009)
  [arXiv:0709.3269 [hep-th]].
  %%CITATION = ARXIV:0709.3269;%%

%\cite{Cossu:2009sq}
\bibitem{Cossu:2009sq}
  G.~Cossu and M.~D'Elia,
  ``Finite size phase transitions in QCD with adjoint fermions,''
  JHEP {\bf 0907}, 048 (2009)
  [arXiv:0904.1353 [hep-lat]].
  %%CITATION = ARXIV:0904.1353;%%

%%%%%%%%% finite temperature %%%%%%%%%%%%%%%%%%

%\cite{Deuzeman:2008sc}
\bibitem{Deuzeman:2008sc}
  A.~Deuzeman, M.~P.~Lombardo and E.~Pallante,
  ``The Physics of eight flavours,''
  Phys.\ Lett.\ B {\bf 670}, 41 (2008)
  [arXiv:0804.2905 [hep-lat]].
  %%CITATION = ARXIV:0804.2905;%%

%\cite{Miura:2011mc}
\bibitem{Miura:2011mc}
  K.~Miura, M.~P.~Lombardo and E.~Pallante,
  ``Chiral phase transition at finite temperature and conformal dynamics in large $N_f$ QCD,''
  Phys.\ Lett.\ B {\bf 710}, 676 (2012)
  [arXiv:1110.3152 [hep-lat]].
  %%CITATION = ARXIV:1110.3152;%%

%\cite{Cheng:2011ic}
\bibitem{Cheng:2011ic}
  A.~Cheng, A.~Hasenfratz and D.~Schaich,
  ``Novel phase in SU(3) lattice gauge theory with 12 light fermions,''
  Phys.\ Rev.\ D {\bf 85}, 094509 (2012)
  [arXiv:1111.2317 [hep-lat]].
  %%CITATION = ARXIV:1111.2317;%%

%\cite{Deuzeman:2012ee}
\bibitem{Deuzeman:2012ee}
  A.~Deuzeman, M.~P.~Lombardo, T.~Nunes Da Silva and E.~Pallante,
  ``The bulk transition of QCD with twelve flavors and the role of improvement,''
  Phys.\ Lett.\ B {\bf 720}, 358 (2013)
  [arXiv:1209.5720 [hep-lat]].
  %%CITATION = ARXIV:1209.5720;%%

%\cite{Miura:2012zqa}
\bibitem{Miura:2012zqa}
  K.~Miura and M.~P.~Lombardo,
    ``Lattice Monte-Carlo study of pre-conformal dynamics in strongly flavoured QCD in the light of the chiral phase transition at finite temperature,''
   Nuclear Physics B {\bf 871}, 52 (2013)
  [arXiv:1212.0955 [hep-lat]].
  %%CITATION = ARXIV:1212.0955;%%
  
%\cite{Kogut:2010cz}
\bibitem{Kogut:2010cz}
  J.~B.~Kogut and D.~K.~Sinclair,
  ``Thermodynamics of lattice QCD with 2 flavours of colour-sextet quarks: A model of walking/conformal Technicolor,''
  Phys.\ Rev.\ D {\bf 81}, 114507 (2010)
  [arXiv:1002.2988 [hep-lat]].
  %%CITATION = ARXIV:1002.2988;%%

%\cite{Kogut:2011ty}
\bibitem{Kogut:2011ty}
  J.~B.~Kogut and D.~K.~Sinclair,
  ``Thermodynamics of lattice QCD with 2 sextet quarks on $N_t$=8 lattices,''
  Phys.\ Rev.\ D {\bf 84}, 074504 (2011)
  [arXiv:1105.3749 [hep-lat]].
  %%CITATION = ARXIV:1105.3749;%%

%\cite{Kogut:2011bd}
\bibitem{Kogut:2011bd}
  J.~B.~Kogut and D.~K.~Sinclair,
  ``Thermodynamics of lattice QCD with 3 flavours of colour-sextet quarks,''
  Phys.\ Rev.\ D {\bf 85}, 054505 (2012)
  [arXiv:1111.3353 [hep-lat]].
  %%CITATION = ARXIV:1111.3353;%%

%\cite{Sinclair:2012fa}
\bibitem{Sinclair:2012fa}
  D.~K.~Sinclair and J.~B.~Kogut,
  ``QCD with colour-sextet quarks,''
  PoS LATTICE {\bf 2012}, 026 (2012)
  [arXiv:1211.0712 [hep-lat]].
  %%CITATION = ARXIV:1211.0712;%%

%\cite{DeGrand:2008kx}
\bibitem{DeGrand:2008kx}
  T.~DeGrand, Y.~Shamir and B.~Svetitsky,
  \ttl{Phase structure of SU(3) gauge theory with two flavors of symmetric-representation fermions,}
  Phys.\ Rev.\  D {\bf 79}, 034501 (2009)
  [arXiv:0812.1427 [hep-lat]].
  %%CITATION = PHRVA,D79,034501;%%

%%%%%%%%% code %%%%%%%%%%%%%%%%%%

\bibitem{Sheikholeslami:1985ij}
  B.~Sheikholeslami and R.~Wohlert,
  \ttl{Improved continuum limit lattice action for QCD with Wilson fermions,}
  Nucl.\ Phys.\  B {\bf 259}, 572 (1985).
  %%CITATION = NUPHA,B259,572;%%

%\cite{Shamir:2010cq}
\bibitem{Shamir:2010cq}
  Y.~Shamir, B.~Svetitsky, and E.~Yurkovsky,
  \ttl{Improvement via hypercubic smearing in triplet and sextet QCD,}
  Phys.\ Rev.\  {\bf D83}, 097502 (2011).
  [arXiv:1012.2819 [hep-lat]].

\bibitem{Hasenfratz:2001hp}
  A.~Hasenfratz and F.~Knechtli,
  \ttl{Flavor symmetry and the static potential with hypercubic blocking,}
  Phys.\ Rev.\  D {\bf 64}, 034504 (2001)
  [arXiv:hep-lat/0103029].
  %%CITATION = PHRVA,D64,034504;%%

%\cite{Hasenfratz:2007rf}
\bibitem{Hasenfratz:2007rf}
  A.~Hasenfratz, R.~Hoffmann and S.~Schaefer,
  \ttl{Hypercubic smeared links for dynamical fermions,}
  JHEP {\bf 0705}, 029 (2007)
  [arXiv:hep-lat/0702028].
%%CITATION = JHEPA,0705,029;%%

%%%%%%%%% phase diagram %%%%%%%%%%%%%%%%%%

%\cite{Iwasaki:1991mr}
\bibitem{Iwasaki:1991mr}
  Y.~Iwasaki, K.~Kanaya, S.~Sakai and T.~Yoshi\'e,
  ``Quark confinement and number of flavors in strong coupling lattice QCD,''
  Phys.\ Rev.\ Lett.\  {\bf 69}, 21 (1992).
  %%CITATION = PRLTA,69,21;%%

%\cite{Iwasaki:2003de}
\bibitem{Iwasaki:2003de}
  Y.~Iwasaki, K.~Kanaya, S.~Kaya, S.~Sakai and T.~Yoshi\'e,
  ``Phase structure of lattice QCD for general number of flavors,''
  Phys.\ Rev.\  D {\bf 69}, 014507 (2004)
  [arXiv:hep-lat/0309159].
  %%CITATION = PHRVA,D69,014507;%%

%\cite{Nagai:2009ip}
\bibitem{Nagai:2009ip}
  K.~Nagai, G.~Carrillo-Ruiz, G.~Koleva and R.~Lewis,
  ``Exploration of SU($N_c$) gauge theory with many Wilson fermions at strong coupling,''
  Phys.\ Rev.\  D {\bf 80}, 074508 (2009)
  [arXiv:0908.0166 [hep-lat]].
  %%CITATION = PHRVA,D80,074508;%%


%\cite{Svetitsky:2010zd}
\bibitem{Svetitsky:2010zd}
  B.~Svetitsky, Y.~Shamir and T.~DeGrand,
  ``Sextet QCD: slow running and the mass anomalous dimension,''
  PoS LATTICE {\bf 2010}, 072 (2010)
  [arXiv:1010.3396 [hep-lat]].
  %%CITATION = ARXIV:1010.3396;%%

%\cite{Svetitsky:2013px}
\bibitem{Svetitsky:2013px}
  B.~Svetitsky,
  ``Conformal or confining---results from lattice gauge theory for higher-representation gauge theories,''
  PoS ConfinementX 271 (2013)
  [arXiv:1301.1877 [hep-lat]].
  %%CITATION = ARXIV:1301.1877;%%

%%%%%%%%% numerics %%%%%%%%%%%%%%%%%%

%\cite{Hasenbusch:2001ne}
\bibitem{Hasenbusch:2001ne}
  M.~Hasenbusch,
  \ttl{Speeding up the Hybrid-Monte-Carlo algorithm for dynamical fermions,}
  Phys.\ Lett.\  B {\bf 519}, 177 (2001)
  [arXiv:hep-lat/0107019].
  %%CITATION = PHLTA,B519,177;%%

%\cite{Urbach:2005ji}
\bibitem{Urbach:2005ji}
 C.~Urbach, K.~Jansen, A.~Shindler and U.~Wenger,
 \ttl{HMC algorithm with multiple time scale integration and mass preconditioning,}
 Comput.\ Phys.\ Commun.\  {\bf 174}, 87 (2006)
 [arXiv:hep-lat/0506011].
 %%CITATION = CPHCB,174,87;%%

%\cite{Takaishi:2005tz}
\bibitem{Takaishi:2005tz}
  T.~Takaishi and P.~de Forcrand,
  \ttl{Testing and tuning new symplectic integrators for hybrid Monte Carlo algorithm in lattice QCD,}
  Phys.\ Rev.\  E {\bf 73}, 036706 (2006)
  [arXiv:hep-lat/0505020].
  %%CITATION = PHRVA,E73,036706;%%

\bibitem{MILC} {\tt http://www.physics.utah.edu/$\sim$detar/milc/}

\end{thebibliography}
\end{document}